\documentclass{article}
\usepackage[latin9]{inputenc}
\usepackage{geometry}
\usepackage{float}
\usepackage{textcomp}
\usepackage{amsmath}
\usepackage{graphicx}

\makeatletter
\@ifundefined{date}{}{\date{}}
\makeatother

\begin{document}
\title{Electromagnetic instability and Schwinger effect in the Witten-Sakai-Sugimoto
model with D0-D4 background}
\author{Wenhe Cai\thanks{whcai@shu.edu.cn} $^{,3,4,5)}$ $\ $Kang-le Li\thanks{lkl96@mail.ustc.edu.cn}
$^{,4)}$ $\ $Si-wen Li \thanks{siwenli@dlmu.edu.cn} $^{,1,2,4)}$}
\maketitle
\begin{center}
$^{1}$Department of Physics, Dalian Maritime University, Dalian 116033,
China
\par\end{center}

\begin{center}
$^{2}$Department of Physics, Center for Field Theory and Particle
Physics, Fudan University, Shanghai 200433, China
\par\end{center}

\begin{center}
$^{3}$Department of Physics, Shanghai University, Shanghai 200444,
China
\par\end{center}

\begin{center}
$^{4}$Department of Modern Physics, University of Science and Technology
of China, Hefei 230026, Anhui, China
\par\end{center}

\begin{center}
$^{5}$Interdisciplinary Center for Theoretical Study, University
of Science and Technology of China, Hefei 230026, Anhui, China
\par\end{center}

\vspace{8mm}

\begin{abstract}
Using the Witten-Sakai-Sugimoto model in the D0-D4 background, we
holographically compute the vacuum decay rate of the Schwinger effect
in this model. Our calculation contains the influence of the D0-brane
density which could be identified as the $\theta$ angle or chiral
potential in QCD. Under the strong electromagnetic fields, the instability
appears due to the creation of quark-antiquark pairs and the associated
decay rate can be obtained by evaluating the imaginary part of the
effective Euler-Heisenberg action which is identified as the action
of the probe brane with a constant electromagnetic field. In the bubble
D0-D4 configuration, we find the decay rate decreases when the $\theta$
angle increases since the vacuum becomes heavier in the present of
the glue condensate in this system. And the decay rate matches to
the result in the black D0-D4 configuration at zero temperature limit
according to our calculations. In this sense, the Hawking-Page transition
of this model could be consistently interpreted as the confined/deconfined
phase transition. Additionally there is another instability from the
D0-brane itself in this system and we suggest that this instability
reflects to the vacuum decay triggered by the $\theta$ angle as it
is known in the $\theta$-dependent QCD.
\end{abstract}
\newpage{}

\tableofcontents{}

\section{Introduction}

Recent years, there have been many advances in the researches on the
strong electromagnetic field, especially in heavy-ion collision since
it is expected that an extremely strong magnetic field is generated
by the collision of the charged particles. In particular, the Schwinger
effect should be one of the most interesting phenomena in the heavy-ion
collision, because the pair creation of charged particles from the
vacuum occurs under such an externally strong electromagnetic field.
In the Schwinger effect, the creation rate of a pair of charged particles
could be obtained by evaluating the imaginary part of Euler-Heisenberg
Lagrangian \cite{key-01,key-02}. However, the result implies the
Schwinger effect is a non-perturbative effect which shows up only
under the strong electromagnetic field.

Although it is still challenging to evaluate the Schwinger effect
under the electromagnetic field, the framework of gauge/gravity duality
or AdS/CFT provides a powerful tool on studying the strongly coupled
quantum field theories \cite{key-03,key-04,key-05}. It has been recognized
that a (d+2)-dimensional classical gravity theory could correspond
to a (d+1)-dimensional gauge theory as a weak/strong duality. So with
this framework, various applications of studying the Schwinger effect
holographically have been presented \cite{key-05+1,key-06,key-07,key-08,key-09,key-010,key-011,key-012,key-013,key-014}.
Particularly, in some top-down holographic approaches e.g. the D3/D7
approach, since the dynamics of the flavors is described by the action
of the probe flavor brane, the associated decay rate to the Schwinger
effect could be evaluated by using the flavored action. Hence it implies
this action could be identified as the holographic Euler-Heisenberg
action and the creation rate of flavored quark-antiquark pairs (i.e.
the vacuum decay rate) can be computed by the imaginary part of this
action \cite{key-012,key-013,key-014}. While this is a different
method, it allows us to quantitatively explore the electromagnetic
instability and Schwinger effect in holography.

On the other hand, the $\theta$-dependence in QCD or Yang-Mills theory
is also interesting \cite{key-015+2,key-015+3}. The $\theta$-dependent
gauge theories contain a Chern-Simons term as a topological term additional
to the action and the coupling of the Chern-Simons term is named as
the $\theta$ angle (as the following form),

\begin{equation}
S=\frac{1}{2g_{YM}^{2}}\int\mathrm{Tr}F\wedge^{*}F-i\frac{\theta}{16\pi^{2}}\int\mathrm{Tr}F\wedge F,\label{eq:1}
\end{equation}
where $g_{YM}$ is the Yang-Mills coupling constant and $F$ is the
gauge field strength. While the experimental value of $\theta$ is
small, the Chern-Simons term leads to many observable phenomena such
as chiral anomaly \cite{key-015+4}, chiral magnetic effect \cite{key-015+5},
deconfinement transition \cite{key-015+6,key-015+7} and effects of
gluon condensate. Accordingly, to investigate the Schwinger effect
with $\theta$-dependence in QCD would be significant since the creation
rate of quark-antiquark pairs is affected by the $\theta$ angle.
So in this paper, we are motivated to study the electromagnetic instability
and the Schwinger effect with such a topological term in holography.

In the gauge/gravity duality, the $\theta$-dependence could be introduced
as the D-brane with D-instanton configuration in the string theory
\cite{key-015,key-015+1}. Thus we use the Witten-Sakai-Sugimoto model
in the D0-D4 background (i.e. the D0-D4/D8 system) where D0-brane
could be the D-instanton in our investigation since this system is
holographically dual to QCD with a Chern-Simons term \cite{key-016,key-017,key-017+1}
(See more details and applications in \cite{key-018,key-019,key-020,key-021}).
In this model, the background geometry is produced by $N_{c}$ coincident
D4-branes wrapped on a cycle with $N_{0}$ smeared D0-branes inside
their worldvolume. The supersymmetry is broken down by imposing the
anti-periodic boundary condition on fermions. In the presence of the
D0-branes, the effective action of the D4-branes takes the following
form,

\begin{eqnarray}
S_{D_{4}} & = & -\mu_{4}\mathrm{Tr}\int d^{4}xdx^{4}e^{-\phi}\sqrt{-\det\left(\mathcal{G}+\mathcal{F}\right)}\nonumber \\
 &  & +\mu_{4}\int C_{5}+\frac{1}{2}\mu_{4}\int C_{1}\wedge\mathcal{F}\wedge\mathcal{F},\label{eq:2}
\end{eqnarray}
where $\mu_{4}=\left(2\pi\right)^{-4}l_{s}^{-5}$, $l_{s}$ is the
length of the string, $\mathcal{G}$ is the induced metric on the
worldvolume. $\mathcal{F}=2\pi\alpha^{\prime}F$ is the gauge field
strength on the D4-brane. $C_{5},\ C_{1}$ is the Romand-Romand 5-
and 1- form respectively. We have used $x^{4}$ to represent the wrapped
direction which is periodic. The first term in (\ref{eq:2}) is the
Dirac-Born-Infeld (DBI) action and the Yang-Mills action comes from
its leading-order expansion respected to $\mathcal{F}$. In the bubble
D0-D4 solution, we have $C_{1}\sim\theta dx^{4}$ \cite{key-017},
thus D0-branes are actually D-instantons and the last term in (\ref{eq:2})
could be integrated as,

\begin{equation}
\int_{S^{1}}C_{1}\sim\theta,\ \ \ \ \int_{S^{1}\times\mathrm{R}^{4}}C_{1}\wedge\mathcal{F}\wedge\mathcal{F}\sim\theta\int_{\mathrm{R}^{4}}\mathcal{F}\wedge\mathcal{F}.\label{eq:3}
\end{equation}

Therefore the number density of D0-branes (D0 charge) is related to
the $\theta$ angle and we could finally obtain the Yang-Mills plus
Chern-Simons action (\ref{eq:1}) from (\ref{eq:2}) and (\ref{eq:3})
as the low-energy theory of the bubble D0-D4 system. As the analysis
of the D4/D8 model (the original Witten-Sakai-Sugimoto model) \cite{key-022,key-022+1,key-023,key-024,key-025},
the bubble D0-D4 corresponds to the confinement phase of the dual
field theory while the black D0-D4 corresponds to the deconfinement
phase\footnote{In \cite{key-026}, an alternative solution for the deconfinement
phase has been proposed .}. However in the black D0-D4 configuration, the physical interpretation
of D0-brane is less clear since D0-brane is not D-instanton. Nevertheless,
we might identify the D0 charge as the chiral potential in the black
D0-D4 configuration according to the phenomenal evidences presented
in \cite{key-027,key-028,key-029}. Besides, the flavors can be introduced
by a stack of $N_{f}$ D8 and anti-D8 branes ($\mathrm{D}8/\overline{\mathrm{D}8}$-branes)
as probes into the D0-D4 background. Hence the chirally symmetric
or broken phase of the dual field theory is represented by the various
configurations of $\mathrm{D}8/\overline{\mathrm{D}8}$-branes in
the bubble or black D0-D4 background.

In this paper, we will focus on the derivation of the effective Euler-Heisenberg
Lagrangian first, then we could explore the electromagnetic instability
and evaluate the creation rate of quark-antiquark pairs in the vacuum.
The paper is organized as follows, in section 2, we review the Witten-Sakai-Sugimoto
model in the D0-D4 background with more details. In section 3, we
derive the the effective Euler-Heisenberg Lagrangian from the probe
$\mathrm{D}8/\overline{\mathrm{D}8}$-branes action in the bubble
and black D0-D4 background respectively. In section 4, we evaluate
the creation rate of quark-antiquark pairs in the vacuum. In the black
D0-D4 background, we find the creation rate is finite at zero temperature
limit which qualitatively coincides with the results from the bubble
D0-D4 background. In this sense, we suggest that the Hawking-Page
transition of this model is suitable to be identified as the confined/deconfined
phase transition. And our numerical calculation shows the creation
rate decreases when the D0 density ($\theta$ angle) increases in
the bubble D0-D4 case. This may be interpreted as that the vacuum
becomes heavier due to the gluon condensate described by this model
in terms of quantum field theory. The final section is the summary
and discussion.

\section{Review of the Witten-Sakai-Sugimoto model in the D0-D4 background}

In this section, let us briefly review the Witten-Sakai-Sugimoto model
in the D0-D4 background (i.e. the D0-D4/D8 system). In string frame,
the background geometry described by the bubble solution of $N_{c}$
D4 branes with smeared $N_{0}$ D0 charges. The near horizon metric
reads in type IIA supergravity \cite{key-017,key-017+1,key-018},

\begin{align}
ds^{2}= & \left(\frac{U}{R}\right)^{3/2}\left[H_{0}^{1/2}\eta_{\mu\nu}dX^{\mu}dX^{\nu}+H_{0}^{-1/2}f\left(U\right)\left(dX^{4}\right)^{2}\right]\nonumber \\
 & +H_{0}^{1/2}\left(\frac{R}{U}\right)^{3/2}\left[\frac{dU^{2}}{f\left(U\right)}+U^{2}d\Omega_{4}^{2}\right].\label{eq:4}
\end{align}
The D0-branes are smeared in the $X^{i},\ i=1,2,3$ and $X^{4}$.
$N_{c}$ represents the number of colors. The dilaton, the field strength
of the Ramond-Ramond field, the function $f(U),\ H_{0}\left(U\right)$
and the radius $R$ of the bulk are given as follows, 
\begin{align}
 & e^{\phi}=g_{s}\left(\frac{U}{R}\right)^{3/4}H_{0}^{3/4},\ \ F_{2}=\frac{1}{\sqrt{2!}}\frac{\mathcal{A}}{U^{4}}\frac{1}{H_{0}^{2}}dU\wedge dX^{4},\ \ F_{4}=\frac{1}{\sqrt{4!}}\mathcal{B}\epsilon_{4},\ \ f(U)\equiv1-\frac{U_{KK}^{3}}{U^{3}},\nonumber \\
 & R^{3}\equiv\pi g_{s}N_{c}l_{s}^{3},\ \ H_{0}=1+\frac{U_{Q_{0}}^{3}}{U^{3}},\ \ \mathcal{A}=\frac{\left(2\pi l_{s}\right)^{7}g_{s}N_{0}}{\omega_{4}V_{4}},\ \ \mathcal{B}=\frac{\left(2\pi l_{s}\right)^{3}g_{s}N_{c}}{\omega_{4}}.\label{eq:5}
\end{align}
where $\mathcal{A},\ \mathcal{B}$ are two integration constants given
as, 
\begin{equation}
\mathcal{A}=3\sqrt{U_{Q_{0}}^{3}\left(U_{Q_{0}}^{3}+U_{KK}^{3}\right)},\ \ \mathcal{B}=3\sqrt{U_{Q_{4}}^{3}\left(U_{Q_{4}}^{3}+U_{KK}^{3}\right)}.
\end{equation}
We use $U_{KK}$, $g_{s}$, $l_{s}$, $V_{4}$ and $\epsilon_{4}$
to represent the coordinate radius of the bottom of the bubble, the
string coupling, the string length, the volume of the unit four sphere
$S^{4}$ and the volume form of the $S^{4}$ respectively. $\alpha^{\prime}$
is defined as $l_{s}^{2}=\alpha^{\prime}$. The coordinate $U$ is
the holographic radial direction, and $U\rightarrow\infty$ corresponds
to the boundary of the bulk. Therefore coordinate $U$ takes the values
in the region $U_{KK}\le U\le\infty$. In order to avoid a possible
singularity at $U=U_{KK}$, the coordinate $U$ satisfies the following
periodic boundary condition \cite{key-016},

\begin{equation}
X^{4}\sim X^{4}+\delta X^{4},\ \ \delta X^{4}=\frac{4\pi}{3}\frac{R^{3/2}}{U_{KK}^{1/2}}H_{0}^{1/2}\left(U_{KK}\right)=2\pi R.\label{eq:7}
\end{equation}
So the Kaluza-Klein mass parameter is obtained,

\begin{equation}
M_{KK}=\frac{2\pi}{\delta X^{4}}=\frac{3U_{KK}^{1/2}}{2R^{3/2}}H_{0}^{-1/2}\left(U_{KK}\right).\label{eq:8}
\end{equation}
The gauge coupling $g_{YM}$ at the cut-off scale $M_{KK}$ in the
4-dimensional Yang-Mills theory is derived as $g_{YM}^{2}=(2\pi)^{2}g_{s}l_{s}/\delta X^{4}$
from the 5-dimensional D4-brane compactified on $S^{1}$. Thus, according
to the gauge/gravity duality and AdS/CFT dictionary, the relationship
between the parameters $R,U_{KK},g_{s},U_{Q_{0}}$ (in the gravity
side) and the parameters $M_{KK},\lambda,N_{c},H_{0}(U_{KK})$ expressed
in QCD is given as,

\begin{equation}
R^{3}=\frac{1}{2}\frac{\lambda l_{s}^{2}}{M_{KK}},\ \ U_{KK}=\frac{2}{9}\lambda M_{KK}l_{s}^{2}H_{KK},\ \ g_{s}=\frac{1}{2\pi}\frac{\lambda}{M_{KK}N_{c}l_{s}},\label{eq:9}
\end{equation}
where $H_{KK}\equiv H_{0}\left(U_{KK}\right)$ and $\lambda\equiv g_{YM}^{2}N_{c}$
is the 4-dimensional 't Hooft coupling. Since one of the spatial coordinates,
denoted by $X^{4}$, is compactified on $S^{1}$, the fermions (and
other irrelevant fields) would be massive by imposing the anti-periodic
boundary condition on the cycle $S^{1}$, thus they are decoupled
from the low-energy theory. Accordingly, the effective theory consists
of the Yang-Mills fields only.

There is an alternatively allowed solution for the D0-D4/D8 system
which is the black brane solution. This solution could be obtained
by interchanging the coordinate $X^{4}$ and $X^{0}$ in (\ref{eq:4}).
So the metric reads \cite{key-021,key-029},

\begin{align}
ds_{D4}^{2}= & \left(\frac{U}{R}\right)^{3/2}\left[-H_{0}^{-1/2}f_{T}\left(U\right)dt^{2}+H_{0}^{1/2}\delta_{ij}dX^{i}dX^{j}+H_{0}^{1/2}\left(dX^{4}\right)^{2}\right]\nonumber \\
 & +H_{0}^{1/2}\left(\frac{R}{U}\right)^{3/2}\left[\frac{dU^{2}}{f_{T}\left(U\right)}+U^{2}d\Omega_{4}^{2}\right].\label{eq:10}
\end{align}
where the function $f_{T}\left(U\right)$ is given as\footnote{In the black D0-D4 solution, we replace $U_{KK}$ by $U_{T}$ in the
solution.},

\begin{equation}
f_{T}\left(U\right)=1-\frac{U_{T}^{3}}{U^{3}}.
\end{equation}
The solution for the other fields in the black D0-D4 background could
also be obtained after interchanging $X^{4}$ and $X^{0}$ in (\ref{eq:5}).
The metric (\ref{eq:10}) describes a horizon at $U=U_{T}$, thus
it corresponds to a quantum field theory at finite temperature.

\begin{figure}[H]
\begin{centering}
\includegraphics[scale=0.4]{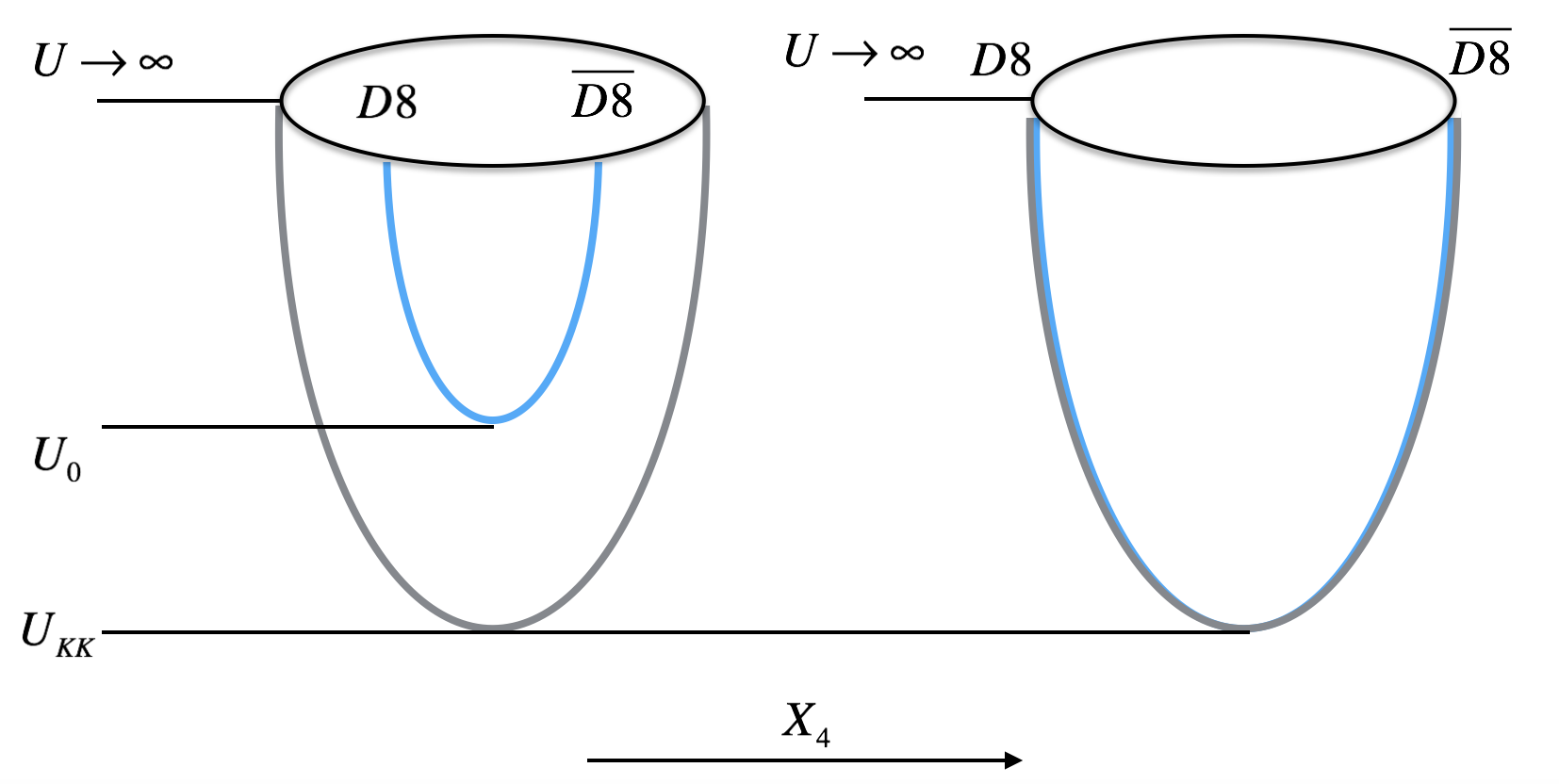} 
\par\end{centering}
\caption{\label{fig:Figure 1}Possible configurations of $\mathrm{D}8/\overline{\mathrm{D}8}$-branes
in the D0-D4 bubble background in $X_{4}-U$ plane. The $\mathrm{D}8/\overline{\mathrm{D}8}$-branes
are always connected which represents the chirally broken phase in
the dual field theory. The left configuration is non-antipodal while
the right one is antipodal. }
\end{figure}

\begin{figure}[H]
\begin{centering}
\includegraphics[scale=0.45]{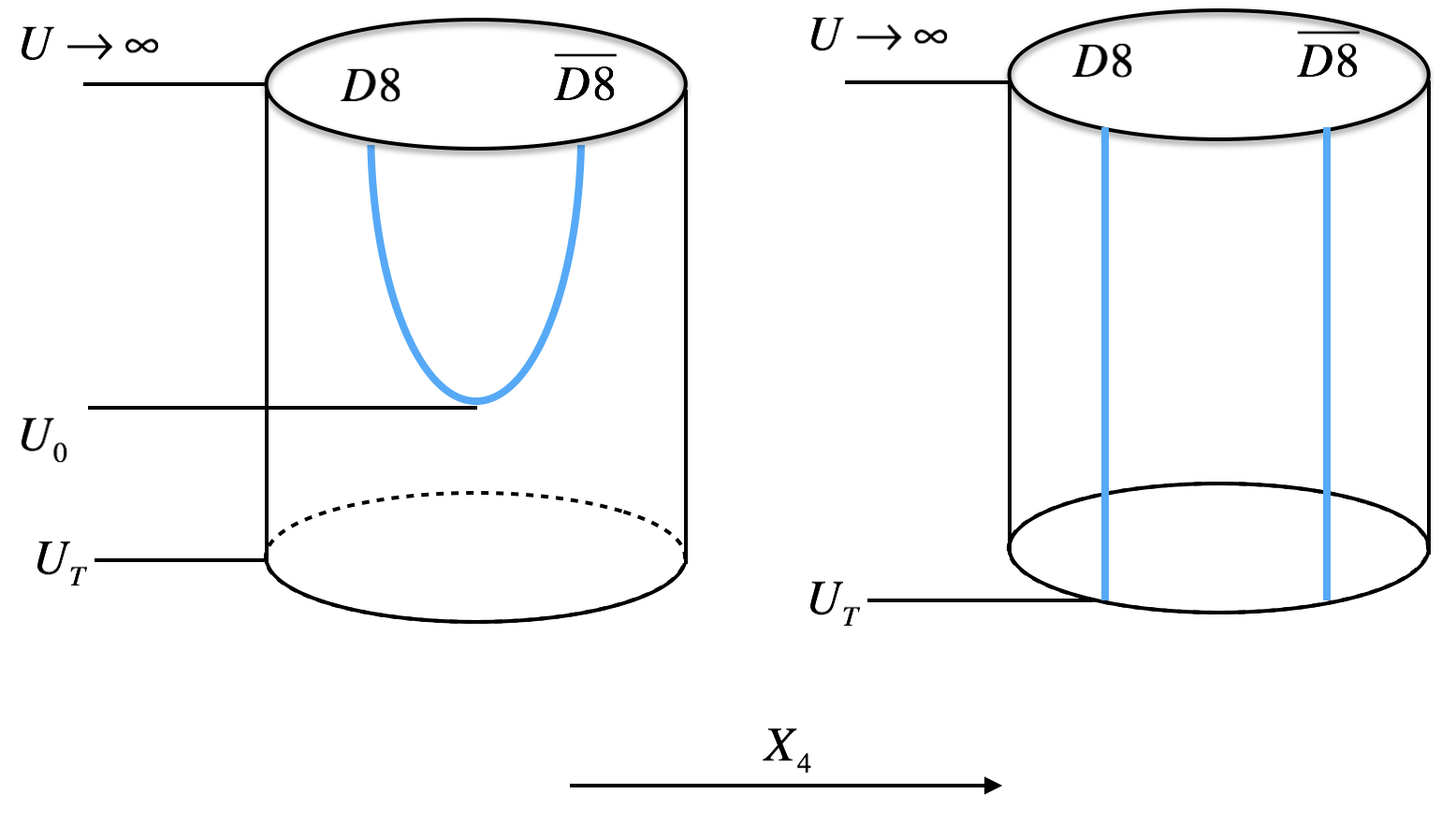} 
\par\end{centering}
\caption{\label{fig:Figure 2}Possible configurations of $\mathrm{D}8/\overline{\mathrm{D}8}$-branes
in the black D0-D4 background in $X_{4}-U$ plane. In the left one,
the $\mathrm{D}8/\overline{\mathrm{D}8}$-branes are connected while
they are parallel in the right one, which corresponds to chirally
broken and symmetric phase in the dual field theory respectively.}
\end{figure}

As the Witten-Sakai-Sugimoto model, the flavors could be introduced
by embedding a stack of $N_{f}$ D8 and anti-D8 branes ($\mathrm{D}8/\overline{\mathrm{D}8}$-branes)
as probes into the D0-D4 background (\ref{eq:4}) or (\ref{eq:10}).
These $\mathrm{D}8/\overline{\mathrm{D}8}$-branes provide $U_{R}\left(N_{f}\right)\times U_{L}\left(N_{f}\right)$
symmetry as chiral symmetry holographically. It has been turned out
that, in the bubble solution (\ref{eq:4}), the $\mathrm{D}8/\overline{\mathrm{D}8}$-branes
are always connected which represents the chirally broken symmetry
in the dual field theory. In the black brane solution (\ref{eq:10}),
the configuration of $\mathrm{D}8/\overline{\mathrm{D}8}$-branes
could be connected or parallel, which could be identified as the chirally
broken or symmetric phase in the dual field theory respectively\footnote{The analyzing of the configuration of the $\mathrm{D}8/\overline{\mathrm{D}8}$-branes
is similar as \cite{key-023,key-024} in the Witten-Sakai-Sugimoto
model. }. The various configurations of $\mathrm{D}8/\overline{\mathrm{D}8}$-branes
are shown in Figure \ref{fig:Figure 1} and Figure \ref{fig:Figure 2}.

\section{Euler-Heisenberg Lagrangian of the D0-D4/D8 brane system}

In this section, we are going to derive the Euler-Heisenberg Lagrangian
in bubble and black D0-D4 brane background respectively. Then we can
investigate the Schwinger effect or creation of quark-antiquark.

\subsection{The bubble D0-D4 geometry}

In the bubble D0-D4 background, the flavored $\mathrm{D}8/\overline{\mathrm{D}8}$-branes
are embedded with the following induced metric on their worldvolume,

\begin{eqnarray}
ds_{D8}^{2} & = & H_{0}^{1/2}\left(\frac{U}{R}\right)^{3/2}\eta_{\mu\nu}dX^{\text{\textmu}}dX^{\nu}+H_{0}^{1/2}\left(\frac{U}{R}\right)^{3/2}\frac{dU^{2}}{h_{c}(U)}\nonumber \\
 & + & H_{0}^{1/2}\left(\frac{R}{U}\right)^{3/2}U^{2}d\Omega_{4}^{2},\label{eq:12}
\end{eqnarray}
where

\begin{align}
h_{c}\left(U\right) & \equiv\left[H_{0}^{-1}f(U)\left(\frac{dX^{4}(U)}{dU}\right)^{2}+\left(\frac{R}{U}\right)^{3}\frac{1}{f(U)}\right]^{-1}.\label{eq:13}
\end{align}
Notice that, for antipodal case, the D8-branes intersect $X^{4}=0$
and the anti-D8-branes put parallel at $X^{4}=\pi R_{S^{1}}$ which
implies $\partial_{U}X^{4}=0$ in (\ref{eq:13}). Here, the $R_{S^{1}}$
represents the radius of $S^{1}$.

In order to evaluate the vacuum decay rate in strongly coupled gauge
theory, we need to derive the Euler-Heisenberg Lagrangian and analyze
its instability (i.e. find the imaginary part of the Euler-Heisenberg
Lagrangian). The Euler-Heisenberg Lagrangian could be obtained from
the probe brane action \cite{key-012,key-013,key-014} since it describes
the dynamics of the flavored quarks. So we shall derive the Euler-Heisenberg
Lagrangian from the following probe $\mathrm{D}8/\overline{\mathrm{D}8}$-brane
action with a constant electromagnetic field\footnote{The low energy effective action of a D-brane consists of two parts:
Dirac-Born-Infeld (DBI) action plus Chern-Simons (CS) term. In this
paper, we do not need to consider the Chern-Simons term since it has
nothing to do with the electromagnetic instability.},

\begin{equation}
S_{D8/\overline{D8}}^{\mathrm{DBI}}=-T_{8}\int_{U_{0}}^{\infty}d^{4}XdUd\Omega_{4}e^{-\phi}\sqrt{-\det\left(P\left[g\right]_{ab}+2\pi\alpha^{\prime}F_{ab}\right)},\label{eq:14}
\end{equation}
where $T_{8}$ is the D8-brane tension which is defined as $T_{8}=g_{s}^{-1}(2\pi)^{-8}l_{s}^{-9}$
and $U_{0}$ represents the connected position of the $\mathrm{D}8/\overline{\mathrm{D}8}$-branes\footnote{For the antipodal case, we need to choose $U_{0}=U_{KK}$.}.
For simplicity, we need to consider a single flavor $N_{f}=1$ for
the presence of an external electromagnetic field. We further require
the electromagnetic field is non-dynamical (i.e. constant) and their
components on the $S^{4}$ are zero. Without losing generality, we
could turn on the electric field on the $X^{1}$ direction only and
the magnetic fields could be introduced in $X^{1},X^{2},X^{3}$ directions
since the $X^{i},\ i=1,2,3$ spacial space is rotationally symmetric.
Inserting constant electromagnetic field with the induced metric (\ref{eq:12})
into the DBI action (\ref{eq:14}), the effective Lagrangian is obtained
as,

\begin{equation}
\mathcal{L}=-\frac{8\pi^{2}T_{8}}{3g_{s}}\int_{U_{0}}^{\infty}dUU^{4}H_{0}^{3/2}h_{c}^{-1/2}\sqrt{\xi_{c}},\label{eq:15}
\end{equation}
where the integral of $d\Omega_{4}$ is Vol($S^{4}$)$=$$8\pi^{2}/3$.
And $\xi_{c}$ is given as, 
\begin{align}
\xi_{c}= & 1-\frac{(2\pi\alpha^{\prime})^{2}R^{3}}{U^{3}H_{0}}\left[F_{01}^{2}-F_{12}^{2}-F_{23}^{2}-F_{13}^{2}+h_{c}(U)(F_{0U}^{2}-F_{1U}^{2})\right]\nonumber \\
 & -\frac{(2\pi\alpha^{\prime})^{4}R^{6}}{U^{6}H_{0}^{2}}\left[F_{01}^{2}F_{23}^{2}+h_{c}(U)\left\{ F_{0U}^{2}(F_{12}^{2}+F_{23}^{2}+F_{13}^{2})-F_{1U}^{2}F_{23}^{2}\right\} \right].
\end{align}
Then we will derive the equations of motion from (\ref{eq:15}) respected
to $A_{0}\left(U\right)$ and $A_{1}\left(U\right)$. Notice that,
we could put $\partial_{i}=0,\ i=1,2,3$ since only the homogeneous
phases are interesting here. In particular, we additionally require
$\partial_{0}=0$ as a constraint of the static configurations, so
that the equations of motion for the static (time-independent) configurations
are obtained as\footnote{When both electric and magnetic fields are turned on, the Chern-Simons
term in the action contributes to the equations of motion. It turns
out that the equations of motion for $A_{U}$ implies the chiral anomaly.
For simplicity, we can ignore this anomaly effect by interpreting
our outcome as the physical values measured at $t=0$, at which $A_{U}$
vanishes as an initial condition.},

\begin{align}
\partial_{U}\left[\frac{UH_{0}^{1/2}h_{c}^{1/2}F_{0U}}{\sqrt{\xi_{c}}}\left\{ 1+\frac{\left(2\pi\alpha^{\prime}\right)^{2}R^{3}}{U^{3}H_{0}}\left(F_{12}^{2}+F_{13}^{2}+F_{23}^{2}\right)\right\} \right] & =0,\nonumber \\
\partial_{U}\left[\frac{UH_{0}^{1/2}h_{c}^{1/2}F_{1U}}{\sqrt{\xi_{c}}}\left\{ 1+\frac{\left(2\pi\alpha^{\prime}\right)^{2}R^{3}F_{23}^{2}}{U^{3}H_{0}}\right\} \right] & =0.
\end{align}
According to the equations of motion, the definition of the number
density $d$ and the current $j$ reads,

\begin{align}
d= & \frac{\left(2\pi\alpha^{\prime}\right)^{2}8\pi^{2}R^{3}T_{8}}{3g_{s}}\frac{UH_{0}^{1/2}h_{c}^{1/2}F_{0U}}{\sqrt{\xi_{c}}}\left\{ 1+\frac{\left(2\pi\alpha^{\prime}\right)^{2}R^{3}}{U^{3}H_{0}}\left(F_{12}^{2}+F_{13}^{2}+F_{23}^{2}\right)\right\} ,\nonumber \\
j= & \frac{\left(2\pi\alpha^{\prime}\right)^{2}8\pi^{2}R^{3}T_{8}}{3g_{s}}\frac{UH_{0}^{1/2}h_{c}^{1/2}F_{1U}}{\sqrt{\xi_{c}}}\left\{ 1+\frac{\left(2\pi\alpha^{\prime}\right)^{2}R^{3}}{U^{3}H_{0}}F_{23}^{2}\right\} .\label{eq:18}
\end{align}
Substituting (\ref{eq:15}) for (\ref{eq:18}), we have

\begin{align}
\xi_{c}= & \frac{1-\frac{\left(2\pi\alpha^{\prime}\right)^{2}R^{3}}{U^{3}H_{0}}\left(F_{01}^{2}-F_{12}^{2}-F_{13}^{2}-F_{23}^{2}\right)-\frac{\left(2\pi\alpha^{\prime}\right)^{4}R^{6}}{U^{6}H_{0}^{2}}F_{01}^{2}F_{23}^{2}}{1+\frac{9g_{s}^{2}}{2^{6}\pi^{4}\left(2\pi\alpha^{\prime}\right)^{2}R^{3}T_{8}^{2}U^{5}H_{0}^{2}}\left[\frac{d^{2}}{1+\frac{\left(2\pi\alpha\prime\right)^{2}R^{3}}{U^{3}H_{0}}\left(F_{12}^{2}+F_{13}^{2}+F_{23}^{2}\right)}-\frac{j^{2}}{1+\frac{\left(2\pi\alpha^{\prime}\right)^{2}R^{3}}{U^{3}H_{0}}F_{23}^{2}}\right]}.\nonumber \\
\end{align}
Therefore the effective Lagrangian as the Euler-Heisenberg Lagrangian
at zero temperature is,

\begin{align}
 & \mathcal{L}=-\frac{8\pi^{2}T_{8}}{3g_{s}}\int_{U_{KK}}^{\infty}dUU^{4}H_{0}^{3/2}h_{c}^{-1/2}\nonumber \\
 & \times\sqrt{\frac{1-\frac{\left(2\pi\alpha^{\prime}\right)^{2}R^{3}}{U^{3}H_{0}}\left(E_{1}^{2}-\vec{B}^{2}\right)-\frac{\left(2\pi\alpha^{\prime}\right)^{4}R^{6}}{U^{6}H_{0}^{2}}E_{1}^{2}B_{1}^{2}}{1+\frac{9g_{s}^{2}}{2^{6}\pi^{4}\left(2\pi\alpha^{\prime}\right)^{2}R^{3}T_{8}^{2}U^{5}H_{0}^{2}}\left[\frac{d^{2}}{1+\frac{\left(2\pi\alpha\prime\right)^{2}R^{3}}{U^{3}H_{0}}\vec{B}^{2}}-\frac{j^{2}}{1+\frac{\left(2\pi\alpha^{\prime}\right)^{2}R^{3}}{U^{3}H_{0}}B_{1}^{2}}\right]}},\label{eq:20}
\end{align}
where we have defined the constant electric and magnetic field as
$F_{0i}=E_{i}$, $\epsilon_{ijk}F_{jk}=B_{i}$ and $\vec{B}^{2}=B_{1}^{2}+B_{2}^{2}+B_{3}^{2}$.

\subsection{The black D0-D4 geometry}

In the black D0-D4 background, there are two possible configurations
for the embedded $\mathrm{D}8/\overline{\mathrm{D}8}$-branes which
are connected and parallel respectively as shown in Figure \ref{fig:Figure 2}.
Generically, the induced metric on the $\mathrm{D}8/\overline{\mathrm{D}8}$-branes
could be written as,

\begin{align}
ds^{2}= & \left(\frac{U}{R}\right)^{3/2}\left[-H_{0}^{-1/2}f_{T}\left(U\right)dt^{2}+H_{0}^{1/2}\delta_{ij}dX^{i}dX^{j}\right]\nonumber \\
 & +\left(\frac{U}{R}\right)^{3/2}H_{0}^{1/2}\frac{dU^{2}}{h_{d}\left(U\right)}+H_{0}^{1/2}\left(\frac{R}{U}\right)^{3/2}U^{2}d\Omega_{4}^{2},\label{eq:21}
\end{align}
where

\begin{equation}
h_{d}\left(U\right)=\left[\left(\partial_{U}X^{4}\right)^{2}+\left(\frac{R}{U}\right)^{3}\frac{1}{f_{T}\left(U\right)}\right]^{-1}.\label{eq:22}
\end{equation}
For connected configuration of the $\mathrm{D}8/\overline{\mathrm{D}8}$-branes,
we keep $X^{4}$ in (\ref{eq:22}) as a generic function depended
on $U$, while $X^{4}$ is a constant for parallel configuration.
Then similar as done in the bubble case, we could obtain the following
Euler-Heisenberg Lagrangian once the induced metic (\ref{eq:21})
and the constant electromagnetic fields are adopted,

\begin{equation}
\mathcal{L}=-\frac{8\pi^{2}T_{8}}{3g_{s}}\int_{U_{0},\ U_{T}}^{\infty}dUU^{4}H_{0}h_{d}^{-1/2}f_{T}^{1/2}\sqrt{\xi_{d}}.\label{eq:23}
\end{equation}
where
\begin{align}
\xi_{d}= & 1-\frac{(2\pi\alpha^{\prime})^{2}R^{3}}{U^{3}H_{0}f_{T}}\left[H_{0}(F_{01}^{2}+F_{0U}^{2}h_{d})-f_{T}(F_{12}^{2}+F_{13}^{2}+F_{23}^{2}+F_{1U}^{2}h_{d})\right]\nonumber \\
 & -\frac{(2\pi\alpha^{\prime})^{4}R^{6}}{U^{6}H_{0}^{2}f_{T}}\left[H_{0}F_{01}^{2}F_{23}^{2}+H_{0}h_{d}F_{0U}^{2}(F_{12}^{2}+F_{13}^{2}+F_{23}^{2})-h_{d}f_{T}F_{1U}^{2}F_{23}^{2}\right].
\end{align}
We have used $U_{0}$ to represent the connected position of the $\mathrm{D}8/\overline{\mathrm{D}8}$-branes.
Notice that in the parallel configuration, the integral in (\ref{eq:23})
starts from $U_{T}$. Next, we could obtain the equations of motion
for static (time independent) configuration which are,

\begin{align}
\partial_{U}\left[\frac{UH_{0}h_{d}^{1/2}F_{0U}}{\sqrt{f_{T}\xi_{d}}}\left\{ 1+\frac{\left(2\pi\alpha^{\prime}\right)^{2}R^{3}}{U^{3}H_{0}}\left(F_{12}^{2}+F_{13}^{2}+F_{23}^{2}\right)\right\} \right] & =0,\nonumber \\
\partial_{U}\left[\frac{Uh_{d}^{1/2}f_{T}^{1/2}F_{1U}}{\sqrt{\xi_{d}}}\left(1+\frac{\left(2\pi\alpha^{\prime}\right)^{2}R^{3}}{U^{3}H_{0}}F_{23}^{2}\right)\right] & =0.
\end{align}
So we have the charge density $d$ and current $j$ defined as, 
\begin{align}
d & =\frac{\left(2\pi\alpha^{\prime}\right)^{2}8\pi^{2}R^{3}T_{8}}{3g_{s}}\frac{UH_{0}h_{d}^{1/2}F_{0U}}{\sqrt{f_{T}\xi_{d}}}\nonumber \\
 & \left\{ 1+\frac{\left(2\pi\alpha^{\prime}\right)^{2}R^{3}}{U^{3}H_{0}}\left(F_{12}^{2}+F_{13}^{2}+F_{23}^{2}\right)\right\} ,\nonumber \\
j & =\frac{\left(2\pi\alpha^{\prime}\right)^{2}8\pi^{2}R^{3}T_{8}}{3g_{s}}\frac{Uh_{d}^{1/2}f_{T}^{1/2}F_{1U}}{\sqrt{\xi_{d}}}\left[1+\frac{\left(2\pi\alpha^{\prime}\right)^{2}R^{3}}{U^{3}H_{0}}F_{23}^{2}\right].
\end{align}
Therefore the on shell effective Lagrangian (\ref{eq:23}) contains
the following form,

\begin{equation}
\xi_{d}=\frac{1-\frac{\left(2\pi\alpha^{\prime}\right)^{2}R^{3}}{U^{3}H_{0}f_{T}}\left[H_{0}F_{01}^{2}-f_{T}\left(F_{12}^{2}+F_{13}^{2}+F_{23}^{2}\right)\right]-\frac{\left(2\pi\alpha^{\prime}\right)^{4}R^{6}}{U^{6}H_{0}f_{T}}F_{01}^{2}F_{23}^{2}}{1+\frac{3^{2}g_{s}^{2}}{2^{6}\pi^{4}\left(2\pi\alpha^{\prime}\right)^{2}R^{3}T_{8}^{2}U^{5}H_{0}^{2}f_{T}}\left[\frac{d^{2}f_{T}}{1+\frac{\left(2\pi\alpha^{\prime}\right)^{2}R^{3}}{U^{3}H_{0}}\left(F_{12}^{2}+F_{13}^{2}+F_{23}^{2}\right)}-\frac{j^{2}H_{0}}{1+\frac{\left(2\pi\alpha^{\prime}\right)^{2}R^{3}}{U^{3}H_{0}}F_{23}^{2}}\right]}.
\end{equation}
Accordingly, the Euler-Heisenberg Lagrangian at finite temperature
is,

\begin{align}
 & \mathcal{L}=-\frac{8\pi^{2}T_{8}}{3g_{s}}\int_{U_{0},\ U_{T}}^{\infty}dUU^{4}H_{0}h_{d}^{-1/2}f_{T}^{1/2}\nonumber \\
 & \times\sqrt{\frac{1-\frac{\left(2\pi\alpha^{\prime}\right)^{2}R^{3}}{U^{3}H_{0}f_{T}}\left[H_{0}E_{1}^{2}-f_{T}\vec{B}^{2}\right]-\frac{\left(2\pi\alpha^{\prime}\right)^{4}R^{6}}{U^{6}H_{0}f_{T}}E_{1}^{2}B_{1}^{2}}{1+\frac{3^{2}g_{s}^{2}}{2^{6}\pi^{4}\left(2\pi\alpha^{\prime}\right)^{2}R^{3}T_{8}^{2}U^{5}H_{0}^{2}f_{T}}\left[\frac{d^{2}f_{T}}{1+\frac{\left(2\pi\alpha^{\prime}\right)^{2}R^{3}}{U^{3}H_{0}}\vec{B}^{2}}-\frac{j^{2}H_{0}}{1+\frac{\left(2\pi\alpha^{\prime}\right)^{2}R^{3}}{U^{3}H_{0}}B_{1}^{2}}\right]}},\label{eq:28}
\end{align}
where the definitions of $E_{i}$ and $B_{i}$ are as same as in the
bubble case.

\section{Holographic pair creation of quark-antiquark}

In the previous sections, we have obtained the effective Euler-Heisenberg
Lagrangian both in the bubble and black D0-D4 brane geometry. In this
section, let us compute the imaginary part of the action in order
to evaluate the pair creation of quark-antiquark holographically at
zero or finite temperature. For simplicity, we will be interested
in looking at the instability of the vacuum as \cite{key-012,key-013,key-014}.

\subsection{Imaginary part of the effective action at zero temperature}

\begin{figure}
\begin{centering}
\includegraphics[scale=0.45]{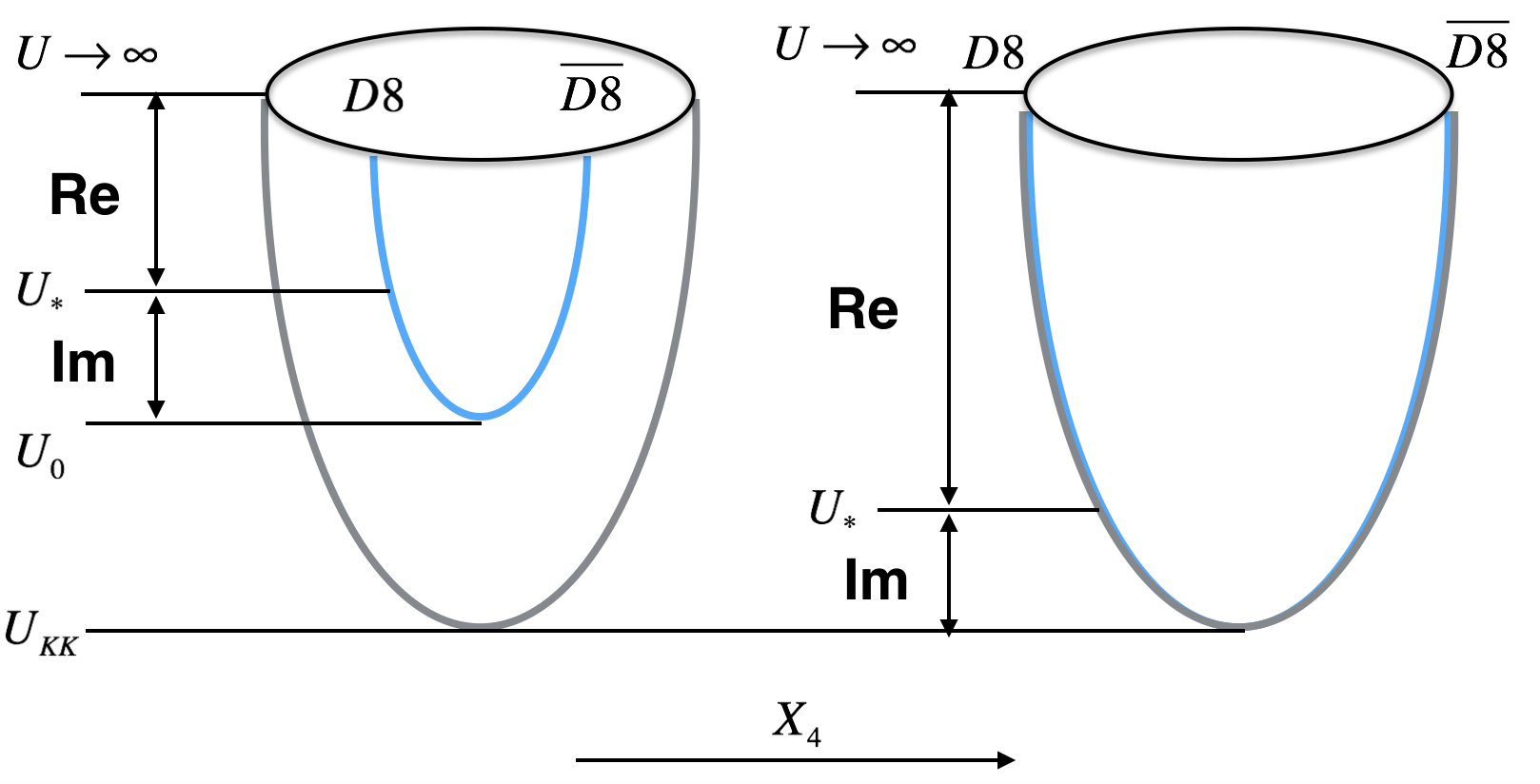} 
\par\end{centering}
\caption{\label{fig:Figure 3}The region of the integral in the imaginary Lagrangian
of the $\mathrm{D}8/\overline{\mathrm{D}8}$-branes in the bubble
D0-D4 background. \textbf{Left}: In the non-antipodal case, the integral
in the imaginary Lagrangian starts from $U_{0}$ to $U_{*}$. \textbf{Right:
}In the antipodal case the integral in the imaginary Lagrangian starts
from $U_{KK}$ to $U_{*}$.}
\end{figure}

Since the bubble D0-D4 geometry holographically corresponds to the
confinement phase of the dual field theory at zero temperature, let
us evaluate the imaginary part of the action (\ref{eq:20}) from the
bubble D0-D4 geometry first. For the vacuum case, by setting $d=j=0$,
the Lagrangian (\ref{eq:20}) takes the following form, 
\begin{equation}
\mathcal{L}=-\frac{8\pi^{2}T_{8}}{3g_{s}}\int_{U_{0},\ U_{KK}}^{\infty}dUU^{4}H_{0}^{3/2}h_{c}^{-1/2}\sqrt{1-\frac{\left(2\pi\alpha^{\prime}\right)^{2}R^{3}}{U^{3}H_{0}}\left(E_{1}^{2}-\vec{B}^{2}\right)-\frac{\left(2\pi\alpha^{\prime}\right)^{4}R^{6}}{U^{6}H_{0}^{2}}E_{1}^{2}B_{1}^{2}}.\label{eq:29}
\end{equation}
The critical position for $U=U_{*}$ , where the imaginary part of
the Lagrangian (\ref{eq:29}) appears, obviously satisfies the equation,

\begin{equation}
\frac{\left(2\pi\alpha^{\prime}\right)^{4}R^{6}}{U_{*}^{6}H_{0}\left(U_{*}\right)^{2}}E_{1}^{2}B_{1}^{2}+\frac{\left(2\pi\alpha^{\prime}\right)^{2}R^{3}}{U_{*}^{3}H_{0}\left(U_{*}\right)}\left(E_{1}^{2}-\vec{B}^{2}\right)-1=0.\label{eq:30}
\end{equation}
After solving (\ref{eq:30}) we therefore obtain, 
\begin{equation}
U_{*}=\left\{ \frac{\left(2\pi\alpha^{\prime}\right)^{2}R^{3}}{2}\left[E_{1}^{2}-\vec{B}^{2}+\sqrt{\left(\vec{B}^{2}-E_{1}^{2}\right)^{2}+4B_{1}^{2}E_{1}^{2}}\right]-U_{Q_{0}}^{3}\right\} ^{1/3}.
\end{equation}
So it is clear that in the region of $U$ : $U_{0}\leq U\leq U_{*}$
, the Lagrangian is imaginary as shown in Figure \ref{fig:Figure 3}.

Since the creation rate of the quark-antiquark is proportional to
the imaginary part of the Lagrangian (\ref{eq:29}), next we are going
to examine whether or not the imaginary part diverges. By the neighborhood
of $U_{0}$, we assume $U_{*}=U_{0}+\varepsilon$ where $\varepsilon\ll U_{0}$
as \cite{key-012}. Then the imaginary part of (\ref{eq:29}) can
be rewritten in terms of the expansion near $U_{0}$ as, 
\begin{align}
\mathrm{Im}\mathcal{L}= & -\frac{8\pi^{2}T_{8}}{3g_{s}}\int_{U_{0},U_{KK}}^{U_{*}}dUU^{4}H_{0}^{3/2}h_{c}^{-1/2}\sqrt{\frac{\left(2\pi\alpha^{\prime}\right)^{4}R^{6}}{U^{6}H_{0}^{2}}E_{1}^{2}B_{1}^{2}+\frac{\left(2\pi\alpha^{\prime}\right)^{2}R^{3}}{U^{3}H_{0}}\left(E_{1}^{2}-\vec{B}^{2}\right)-1}\nonumber \\
\simeq & -\frac{8\pi^{2}T_{8}}{3g_{s}}\mathcal{F}\left(U_{0}\right)\underset{U_{*}\rightarrow U_{0}}{\lim}\left[\left(U_{*}-U_{0}\right)\times h_{c}^{-1/2}\left(U_{0}\right)\right]\nonumber \\
\simeq & -\frac{8\pi^{2}T_{8}}{3g_{s}}\mathcal{F}\left(U_{0}\right)\left[H_{0}^{-1}\left(U_{0}\right)f(U_{0})\left[X^{4}\left(U_{0}+\varepsilon\right)-X^{4}\left(U_{0}\right)\right]^{2}+\frac{R^{3}\varepsilon^{2}}{\left(U_{0}+\varepsilon\right)^{3}-U_{KK}^{3}}\right]^{1/2},\label{eq:32}
\end{align}
where $\mathcal{F}\left(U\right)$ is defined as, 
\begin{equation}
\mathcal{F}(U)=U^{4}H_{0}^{3/2}\sqrt{\frac{(2\pi\alpha^{\prime})^{4}R^{6}}{U^{6}H_{0}^{2}}E_{1}^{2}B_{1}^{2}+\frac{(2\pi\alpha^{\prime})^{2}R^{3}}{U^{3}H_{0}}(E_{1}^{2}-\vec{B}^{2})-1}.
\end{equation}
Consequently (\ref{eq:32}) is finite for the non-antipodal case (i.e.
$U_{0}>U_{KK}$). Moreover, in the antipodal case, we must require
$U_{0}=U_{KK}$. So $X^{4}\left(U_{0}+\varepsilon\right)-X^{4}\left(U_{0}\right)$
vanishes since $X^{4}$ is a constant. Thus (\ref{eq:32}) could be
simplified as,

\begin{align}
\mathrm{Im}\mathcal{L}\simeq & \frac{8\pi^{2}T_{8}R^{3/2}}{3g_{s}}\mathcal{F}\left(U_{KK}\right)\frac{\varepsilon}{\sqrt{\left(U_{KK}+\varepsilon\right)^{3}-U_{KK}^{3}}}\nonumber \\
\simeq & \frac{8\pi^{2}T_{8}R^{3/2}}{3g_{s}U_{KK}}\mathcal{F}\left(U_{KK}\right)\left[\sqrt{\frac{\varepsilon}{3}}+\mathcal{O}\left(\varepsilon^{3/2}\right)\right]=\mathrm{finite}.
\end{align}
So (\ref{eq:32}) shows a finite value for the creation rate of quark-antiquark
in our D0-D4/D8 system which is due to the confining scale $U_{KK}$.
If setting $U_{Q_{0}}=0$ (i.e. no smeared D0-branes or vanished $\theta$
angle), our result returns to the approximated approach in the Witten-Sakai-Sugimoto
model as \cite{key-012}. Accordingly, it is natural to treat (\ref{eq:32})
as a holographically generic form of the creation rate of quark-antiquark
which is dependent on the topological charge from the QCD vacuum.

Since the creation of the quark antiquark breaks the vacuum, we need
to evaluate the critical electric field. By the condition that the
Lagrangian (\ref{eq:32}) begins to be imaginary, the critical electric
field could be derived by solving, 
\begin{equation}
U_{0}=U_{*}=\left\{ \frac{\left(2\pi\alpha^{\prime}\right)^{2}R^{3}}{2}\left[E_{1}^{2}-\vec{B}^{2}+\sqrt{\left(\vec{B}^{2}-E_{1}^{2}\right)^{2}+4B_{1}^{2}E_{1}^{2}}\right]-U_{Q_{0}}^{3}\right\} ^{1/3}.
\end{equation}
Thus the critical electric field is,

\begin{equation}
E_{cr}=\left[\frac{U_{0}^{3}+U_{Q_{0}}^{3}}{\left(2\pi\alpha^{\prime}\right)^{2}R^{3}}\frac{\left\{ \frac{U_{0}^{3}+U_{Q_{0}}^{3}}{\left(2\pi\alpha^{\prime}\right)^{2}R^{3}}+\vec{B}^{2}\right\} }{\left\{ \frac{U_{0}^{3}+U_{Q_{0}}^{3}}{\left(2\pi\alpha^{\prime}\right)^{2}R^{3}}+B_{1}^{2}\right\} }\right]^{1/2}.\label{eq:36}
\end{equation}
We find the critical electric field does not depend on $B_{1}$ if
setting $B_{2},B_{3}=0$. Besides, if we look at the antipodal case
which means $U_{0}=U_{KK}$, and substitute (\ref{eq:36}) for (\ref{eq:9}),
then the critical electric field could be obtained as,

\begin{equation}
E_{cr}=\frac{2}{27\pi}\lambda M_{KK}^{2}\left(1+\zeta\right)^{2}\left[\frac{\frac{4}{3^{6}\pi^{2}}\lambda^{2}M_{KK}^{4}\left(1+\zeta\right)^{4}+\vec{B}^{2}}{\frac{4}{3^{6}\pi^{2}}\lambda^{2}M_{KK}^{4}\left(1+\zeta\right)^{4}+B_{1}^{2}}\right]^{1/2}.\label{eq:37}
\end{equation}
where $\zeta$ is defined as $\zeta=U_{Q_{0}}^{3}/U_{KK}^{3}$. Since
$\zeta$ is related to the D0-brane density, (\ref{eq:37}) shows
the dependence on the $\theta$ angle from the QCD vacuum. And it
coincides with the generic formula in \cite{key-012} if setting $\zeta=0$.
Notice that if $B_{2}=B_{3}=0$, the critical electric field is $E_{cr}=\frac{2}{27\pi}\lambda M_{KK}^{2}\left(1+\zeta\right)^{2}$
which increases by the appearance of the $\theta$ angle (i.e. D0
charge).

In order to evaluate the imaginary part of the Lagrangian (\ref{eq:29})
numerically, let us derive the expression of (\ref{eq:29}) by using
the following dimensionless variables. Introducing the dimensionless
variables as,

\begin{equation}
U=\frac{U_{KK}}{yH_{KK}H_{0}^{1/3}}=\frac{\left(U_{KK}^{3}-H_{KK}^{3}U_{Q_{0}}^{3}y^{3}\right)^{1/3}}{yH_{KK}},\label{eq:38}
\end{equation}
and substituting (\ref{eq:29}) for (\ref{eq:9}) and (\ref{eq:38}),
we obtain, 
\begin{align}
\mathrm{Im}\mathcal{L}= & \frac{M_{KK}^{4}\lambda^{3}N_{c}}{2\cdot3^{8}\pi^{5}}\int_{y_{*}}^{y_{0},y_{kk}}dy\left\{ 1+y\left[1-y^{3}\zeta\left(1+\zeta\right)^{3}\right]^{4/3}\left[1-y^{3}\left(1+\zeta\right)^{4}\right]^{2}x_{4}^{\prime2}\right\} ^{1/2}\nonumber \\
 & \times y^{-9/2}\left[1-y^{3}\zeta\left(1+\zeta\right)^{3}\right]^{-5/6}\left[1-y^{3}\left(1+\zeta\right)^{4}\right]^{-1/2}\sqrt{y^{6}\boldsymbol{\mathrm{E}}_{1}^{2}\boldsymbol{\mathrm{B}}_{1}^{2}+y^{3}\left(\boldsymbol{\mathrm{E}}_{1}^{2}-\vec{\boldsymbol{\mathrm{B}}}^{2}\right)-1},\label{eq:39}
\end{align}
where 
\[
y_{kk}=\left(1+\zeta\right)^{-4/3},y_{*}=\frac{U_{KK}}{\left(1+\zeta\right)\left(U_{*}^{3}+U_{Q_{0}}^{3}\right)^{1/3}},\ \ X_{4}=\frac{3x_{4}}{2M_{KK}},\boldsymbol{\mathrm{E}}_{i}=\frac{3^{3}\pi}{2\lambda M_{KK}^{2}}E_{i},\ \boldsymbol{\mathrm{B}}_{i}=\frac{3^{3}\pi}{2\lambda M_{KK}^{2}}B_{i}.
\]
Notice that we need to obtain the exact formula for $x_{4}$ in (\ref{eq:39})
by solving its equation of motion before the numerical calculations,
however, which would become very challenging. So in this paper, we
will not attempt to evaluate (\ref{eq:39}) with the exact solution
for $x_{4}$. Instead, as a typical exploration, we evaluate the imaginary
part (\ref{eq:29}) in the antipodal case for simplicity\footnote{In some limit, the contribution from $x_{4}^{\prime}$ to the effective
Lagrangian is not important. For example, if $\varepsilon\rightarrow0$
in (\ref{eq:32}), we obtain $X^{4}\left(U_{0}+\varepsilon\right)\simeq X^{4}\left(U_{0}\right)$.
Therefore in this limit, $x_{4}^{\prime}$ does not contribute to
the effective Lagrangian (\ref{eq:32}).}. Moreover, we interestingly find that (\ref{eq:39}) may show an
additionally possible instability because the second line of (\ref{eq:39})
could also be imaginary and it does not depend on the electromagnetic
field. Since the bubble D0-D4/D8 system corresponds to a confining
Yang-Mills theory with a topological Chern-Simons term, so the instability
produced by $\theta$ angle (i.e. D0 charge) could be holographically
interpreted as the transition between the different $\theta$ vacuum
states, which has been very well-known in QCD. However in order to
investigate the electromagnetic instability, we need to remove the
$\theta$-instability from the vacuum. Hence we expand (\ref{eq:39})
by $\zeta$ since $\theta$ angle in QCD is very small. Therefore
in the antipodal case (i.e. $x_{4}^{\prime}=0$) with small $\zeta$
expansion, we obtain the following formula for the imaginary Lagrangian,

\begin{align}
\mathrm{Im}\mathcal{L}\simeq & \frac{M_{KK}^{4}\lambda^{3}N_{c}}{2\cdot3^{8}\pi^{5}}\int_{y_{*}}^{y_{kk}}dy\sqrt{y^{6}\boldsymbol{\mathrm{E}}_{1}^{2}\boldsymbol{\mathrm{B}}_{1}^{2}+y^{3}\left(\boldsymbol{\mathrm{E}}_{1}^{2}-\vec{\boldsymbol{\mathrm{B}}}^{2}\right)-1}\nonumber \\
 & \times y^{-9/2}\left[\frac{1}{\sqrt{1-y^{3}}}-\frac{y^{3}\left(5y^{3}-17\right)}{6\left(1-y^{3}\right)^{3/2}}\zeta\right].\label{eq:40}
\end{align}

\begin{figure}[H]
\begin{centering}
\includegraphics[scale=0.3]{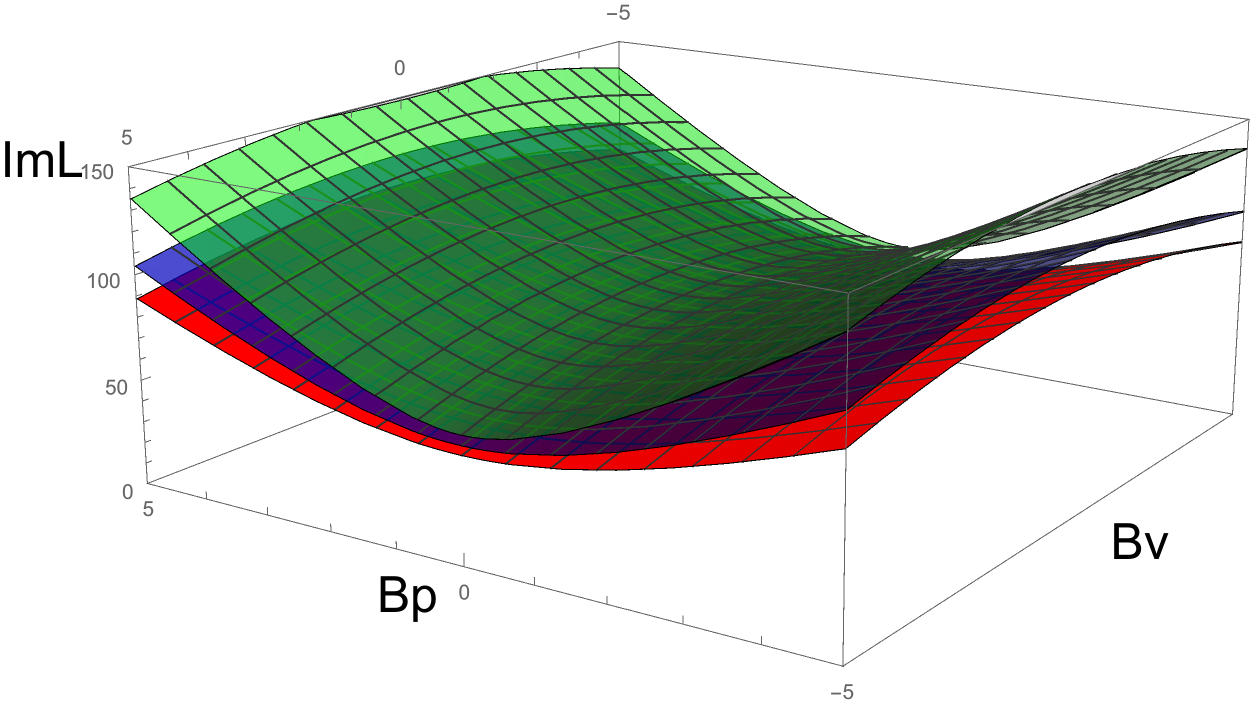} 
\par\end{centering}
\caption{\label{fig:Figure 4}The imaginary part of the effective Lagrangian
with a fixed $E$ as a function of $\zeta$, the magnetic field $B_{\mathrm{P}}$
and $B_{\mathrm{V}}$ which is parallel and perpendicular to the electric
field $E$. In this figure, we look at $\boldsymbol{\mathrm{E}}_{1}=\mathrm{E}=10$
with $\zeta=0.5,\ 0.3,\ 0$ represented by red, blue and green respectively.
It shows the dependence on $\zeta$ ($\theta$ angle) and the imaginary
part of the Lagrangian decreases if $\zeta$ increases.}
\end{figure}

\begin{figure}[H]
\begin{centering}
\includegraphics[scale=0.37]{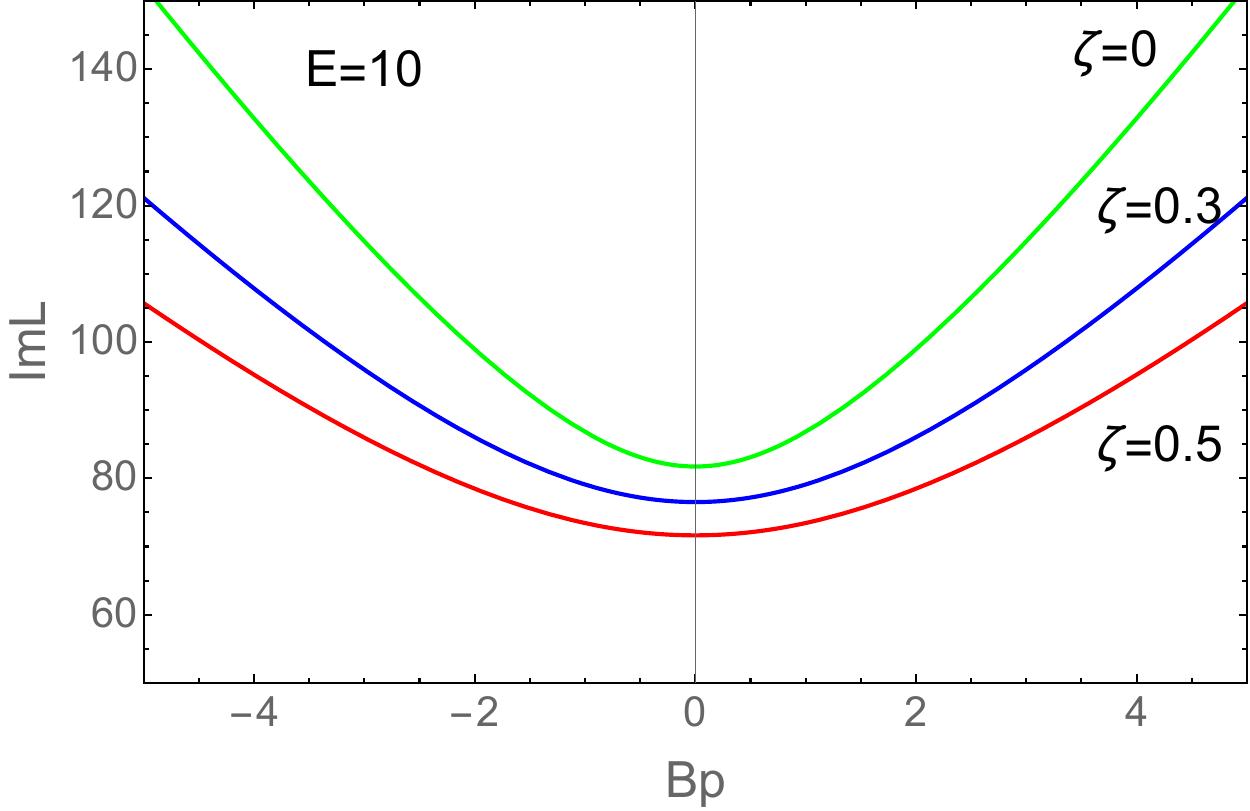}\includegraphics[scale=0.37]{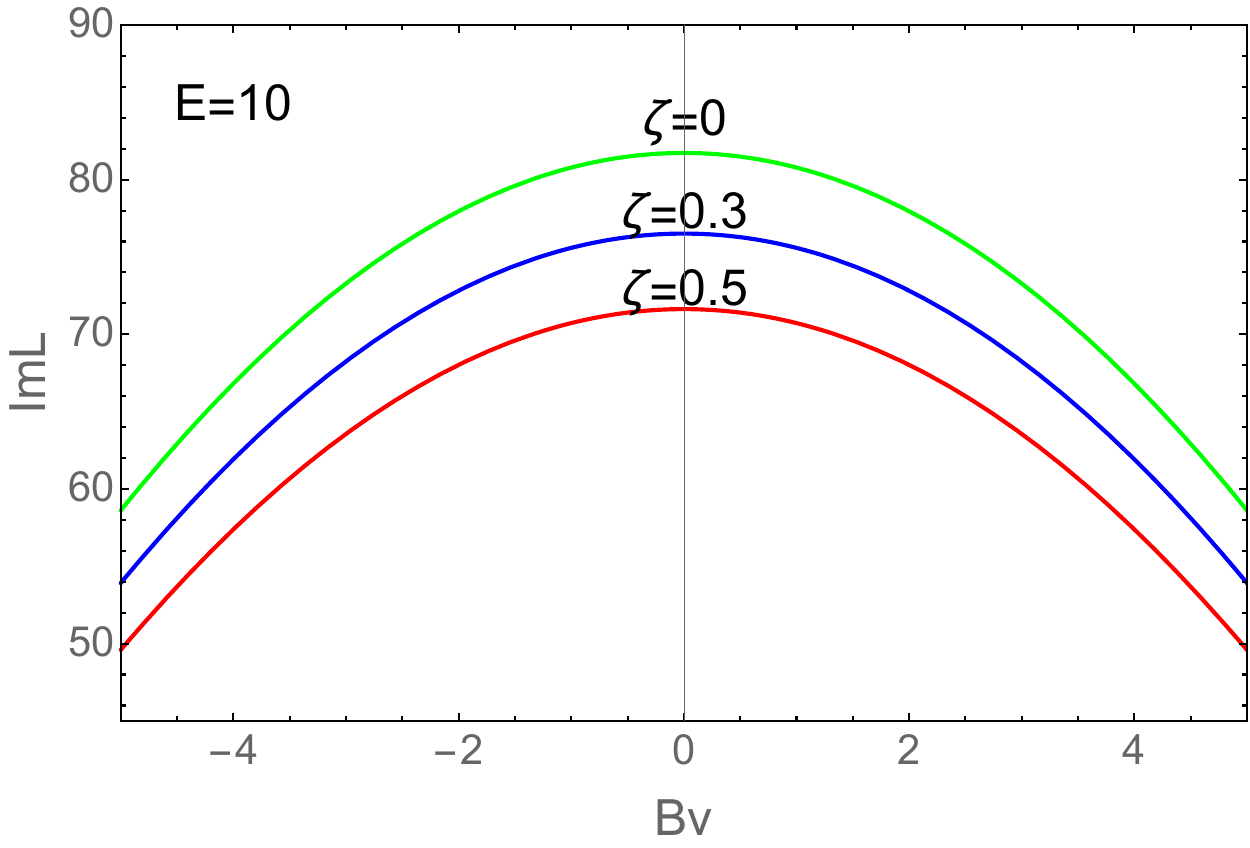} 
\par\end{centering}
\caption{\label{fig:Figure 5}Relation between $\mathrm{Im}L$ and $B_{\mathrm{V}},\ B_{\mathrm{P}}$
with different $\zeta$. If $\zeta$ increases, $\mathrm{Im}L$ decreases.
\textbf{Left:} The magnetic field is parallel to the electric field
and $\mathrm{Im}L$ increases as the magnetic field increases. \textbf{Right:}
The magnetic field is perpendicular to the electric field and $\mathrm{Im}L$
decreases as the magnetic field increases. }
\end{figure}

The (\ref{eq:40}) could be numerically evaluated and the result is
shown in Figure \ref{fig:Figure 4}. We plot the $\mathrm{Im}L$ as
a function of $\zeta$, the dimensionless magnetic field $B_{\mathrm{P}}$
and $B_{\mathrm{V}}$ which is parallel and perpendicular to the (dimensionless)
electric field $\boldsymbol{\mathrm{E}}$. In Figure \ref{fig:Figure 4},
we use different colors, as red, blue and green, to distinguish the
dependence on $\zeta$ ($\theta$ angle) with $\zeta=0.5,\ 0.3,\ 0$
respectively. It shows if the magnetic field and the electric field
are parallel, the imaginary part of the Lagrangian increases as the
magnetic field increases. And on the other hand, if the magnetic field
and the electric field is perpendicular to each other, the imaginary
part of the Lagrangian decreases when the magnetic field increases.
The relation between $\mathrm{Im}L$ and parallel/perpendicular magnetic
field with different $\zeta$ is shown in Figure \ref{fig:Figure 5}.

\begin{figure}[H]
\begin{centering}
\includegraphics[scale=0.37]{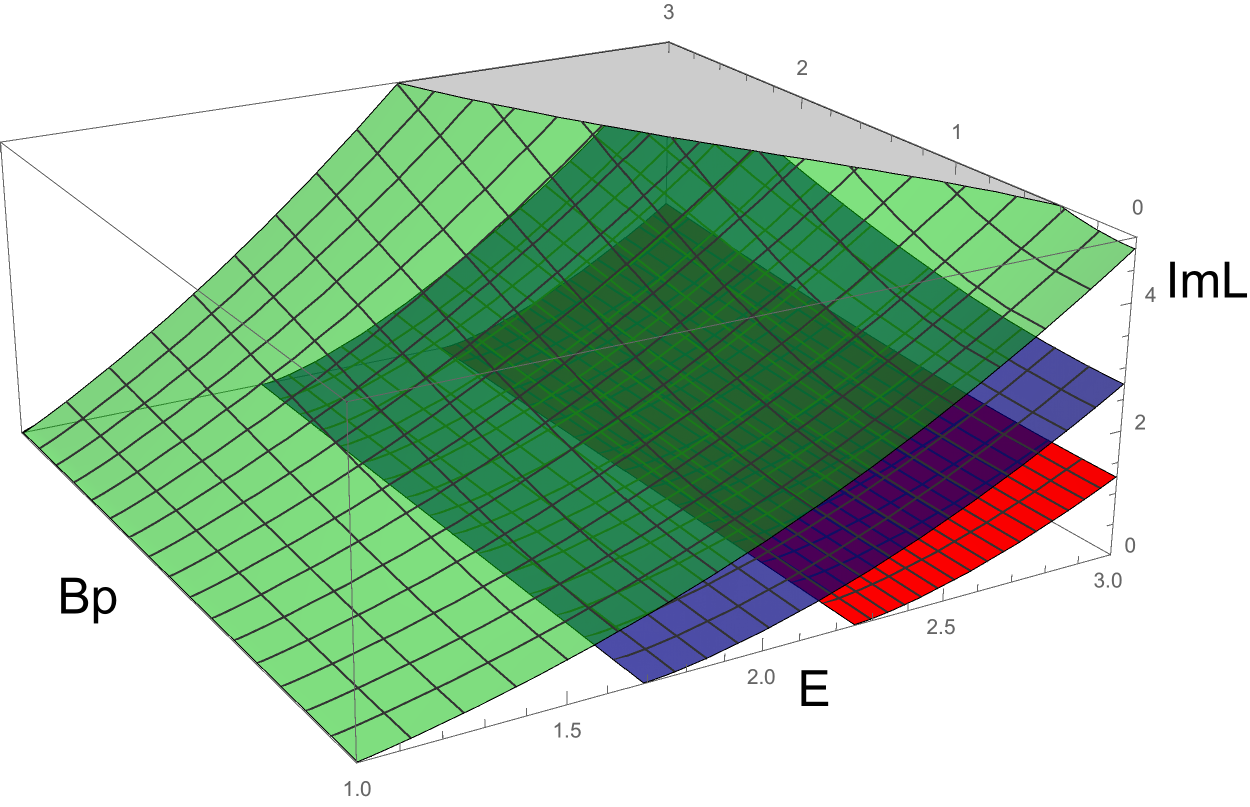}\includegraphics[scale=0.37]{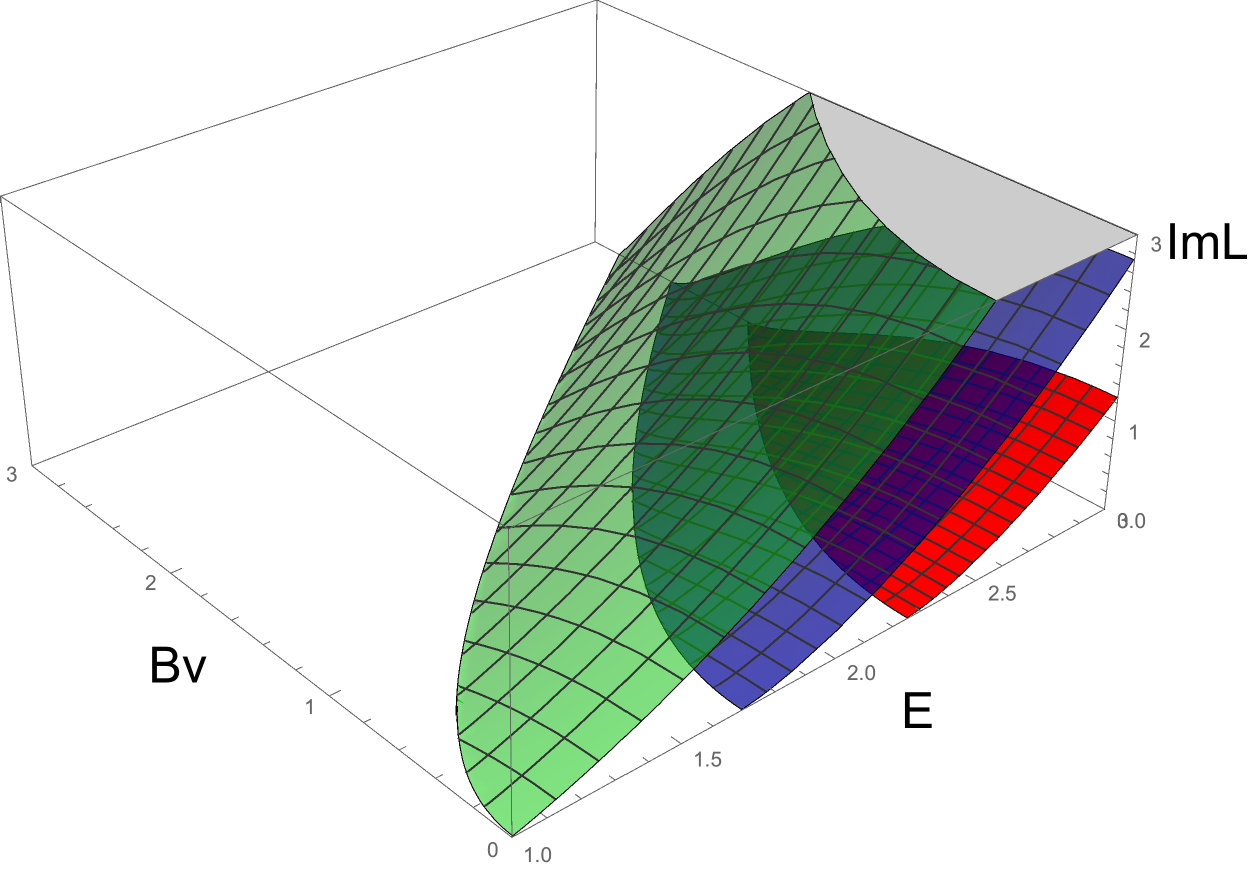} 
\par\end{centering}
\caption{\label{fig:Figure 6}In this figure, red blue and green represent
$\zeta=0.5,\ 0.3,\ 0$ respectively as before. \textbf{Left:} The
magnetic field is parallel to the electric field. \textbf{Right:}
The magnetic field is perpendicular to the electric field.}
\end{figure}

\begin{figure}[H]
\begin{centering}
\includegraphics[scale=0.4]{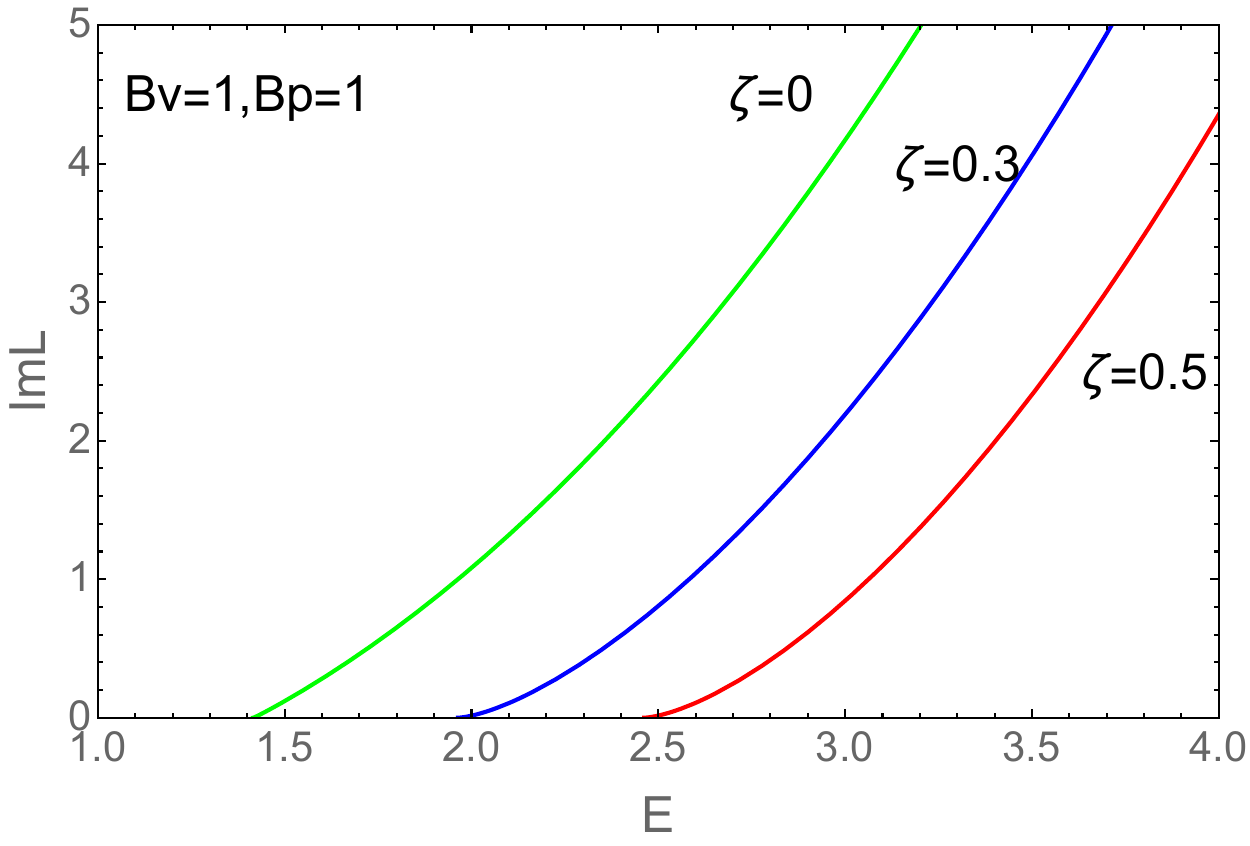} 
\par\end{centering}
\caption{\label{fig:Figure 7}The relation between $\mathrm{Im}L$ with an
arbitrary magnetic field and various $\zeta$.}
\end{figure}

We furthermore look at the dependence on $E$ with various $\zeta$
in the case of a parallel/perpendicular magnetic field respectively.
The numerical evaluation is summarized in Figure \ref{fig:Figure 6}.
Accordingly, we could conclude that, in the bubble D0-D4 system, the
electromagnetic instability is suppressed by the appearance of D0
charge ($\theta$ angle in QCD), however with a fixed $\zeta$, its
behavior depends on the direction of the magnetic field relative to
the electric field. Finally, we also plot the relation between the
electric field with an arbitrary magnetic field to confirm our conclusion
which is shown in Figure \ref{fig:Figure 7}.

\subsection{Imaginary part of the action at finite temperature}

In the previous section, we have obtained the effective action from
the black D0-D4 background i.e. at finite temperature. Since we are
most interested in the vacuum instability, it would be suitable to
set the charge density and current vanished in the Lagrangian (\ref{eq:28})
i.e. $d=j=0$ . Then we obtain the simplified Lagrangian from (\ref{eq:28})
which is,

\begin{align}
\mathcal{L}= & -\frac{8\pi^{2}T_{8}}{3g_{s}}\int_{U_{0},U_{T}}^{\infty}dUU^{4}H_{0}h_{d}^{-1/2}f_{T}^{1/2}\nonumber \\
 & \times\sqrt{1-\frac{\left(2\pi\alpha^{\prime}\right)^{2}R^{3}}{U^{3}H_{0}f_{T}}\left[H_{0}E_{1}^{2}-f_{T}\vec{B}^{2}\right]-\frac{\left(2\pi\alpha^{\prime}\right)^{4}R^{6}}{U^{6}H_{0}f_{T}}E_{1}^{2}B_{1}^{2}}.\label{eq:41}
\end{align}
We can evaluate the imaginary part of the Lagrangian to derive the
rate of the quark antiquark creation in the vacuum by (\ref{eq:41}).

To begin with, let us consider the case of zero-temperature limit,
i.e. $U_{T}\rightarrow0$, so that the function $f_{T}\left(U\right)$
approaches unity. The third term in the square root of the Lagrangian
(\ref{eq:41}) must be dominated because the integral in (\ref{eq:41})
should be totally imaginary. Thus the $U$-integral is finite. Accordingly,
in the presence of the electromagnetic field, the vacuum decay rate
is finite at strong coupling in the zero temperature limit in our
D0-D4/D8 system or the Witten-Sakai-Sugimoto model. Then let us compute
it in details. Introducing the dimensionless variables as,

\begin{equation}
y=\frac{U_{T}}{UH_{T}H_{0}^{1/3}},\ \zeta_{T}=\frac{U_{Q_{0}}^{3}}{U_{T}^{3}},\ \chi=\frac{U_{T}^{3}}{\left(2\pi\alpha^{\prime}\right)^{2}R^{3}},\ x_{4}=\frac{3}{2}\frac{3U_{T}^{1/2}}{2R^{3/2}}H_{T}^{-1/2}X^{4},
\end{equation}
we have, 
\begin{align}
\mathcal{L}= & -\frac{2^{3}\pi^{2}\left(2\pi\alpha^{\prime}\right)^{7/3}R^{5}T_{8}}{3g_{s}}\chi^{7/6}\left(1+\zeta_{T}\right)^{-1/2}\int_{y_{T},y_{0}}^{0}dyy^{-9/2}\left[1-y^{3}\zeta_{T}\left(1+\zeta_{T}\right)^{3}\right]^{-5/6}\nonumber \\
 & \times\left\{ y\left[1-y^{3}\left(1+\zeta_{T}\right)^{4}\right]\left[1-y^{3}\zeta_{T}\left(1+\zeta_{T}\right)^{3}\right]^{4/3}x_{4}^{\prime2}+1\right\} ^{1/2}\nonumber \\
 & \times\sqrt{\left(1+\zeta_{T}\right)^{-6}-\frac{y^{3}}{\chi}\left\{ E_{1}^{2}\left[1-y^{3}\left(1+\zeta_{T}\right)^{4}\right]^{-1}-\vec{B}^{2}\right\} \left(1+\zeta_{T}\right)^{-3}-\frac{y^{6}}{\chi^{2}}E_{1}^{2}B_{1}^{2}\left[1-y^{3}\left(1+\zeta_{T}\right)^{4}\right]^{-1}},\label{eq:43}
\end{align}
where $y_{T}=\left(1+\zeta_{T}\right)^{-4/3}$ and $H_{T}\equiv H_{0}\left(U_{T}\right)$.
Further rescale $Y=\chi^{-1/3}y$, it yields, 
\begin{align}
\mathrm{Im}\mathcal{L}= & -\frac{2^{3}\pi^{2}\left(2\pi\alpha^{\prime}\right)^{7/3}R^{5}T_{8}}{3g_{s}}\left(1+\zeta_{T}\right)^{-7/2}\int_{\chi^{-1/3}\left(1+\zeta_{T}\right)^{-4/3}}^{Y_{*}}dYY^{-9/2}\bigg\{\left[B_{1}^{2}E_{1}^{2}\left(1+\zeta_{T}\right)^{6}+\vec{B}^{2}\left(1+\zeta_{T}\right)^{7}\chi\right]Y^{6}\nonumber \\
 & +\left[\left(1+4\zeta_{T}+6\zeta_{T}^{2}+4\zeta_{T}^{3}+\zeta_{T}^{4}\right)\chi+\left(E_{1}^{2}-\vec{B}^{2}\right)\left(1+\zeta_{T}\right)^{3}\right]Y^{3}-1\bigg\}^{1/2}\nonumber \\
 & \times\bigg\{1+\chi^{1/3}Y\left[1-Y^{3}\zeta_{T}\left(1+\zeta_{T}\right)^{3}\chi\right]^{4/3}\left[1-Y^{3}\left(1+\zeta_{T}\right)^{4}\chi\right]x_{4}^{\prime2}\bigg\}^{1/2}\left[1-Y^{3}\zeta_{T}\left(1+\zeta_{T}\right)^{3}\chi\right]^{-5/6}\nonumber \\
 & \times\left[1-Y^{3}\left(1+\zeta_{T}\right)^{4}\chi\right]^{-1/2}.\label{eq:44}
\end{align}
Since there are two possible configurations for $\mathrm{D}8/\overline{\mathrm{D}8}$-branes
in the black brane background as shown in Figure \ref{fig:Figure 2},
let us consider the parallel $\mathrm{D}8/\overline{\mathrm{D}8}$-branes
first, i.e. $x_{4}^{\prime}=0$. In the small temperature limit $\chi\rightarrow0$,
the third term in the square root of the Lagrangian (\ref{eq:41})
become dominated, so that we have the following behavior from (\ref{eq:44}),
\begin{align}
\mathrm{Im}\mathcal{L}\simeq & -\frac{2^{3}\pi^{2}\left(2\pi\alpha^{\prime}\right)^{7/3}R^{5}T_{8}}{3g_{s}}\vec{E}\cdot\vec{B}\left(1+\zeta_{T}\right)^{-1/2}\int_{\chi^{-1/3}\left(1+\zeta_{T}\right)^{-4/3}}^{Y_{*}}dYY^{-3/2}\nonumber \\
 & \times\left[1-Y^{3}\zeta_{T}\left(1+\zeta_{T}\right)^{3}\chi\right]^{-5/6}\left[1-Y^{3}\zeta_{T}\left(1+\zeta_{T}\right)^{4}\chi\right]^{-1/2}\nonumber \\
\simeq & \frac{2^{4}\pi^{2}\left(2\pi\alpha^{\prime}\right)^{7/3}R^{5}T_{8}}{3g_{s}\sqrt{1+\zeta_{T}}}\vec{B}\cdot\vec{E}Y_{*}^{-1/2}+\mathcal{O}\left(\chi^{1/6}\right)=\mathrm{finite}.\label{eq:45}
\end{align}
where $Y_{*}$ satisfies the following equation in the small $\chi$
limit,

\begin{equation}
1-\left(E_{1}^{2}-\vec{B}^{2}\right)Y_{*}^{3}-\left(B_{1}E_{1}\right)^{2}Y_{*}^{6}=0,\label{eq:46}
\end{equation}
so that if the magnetic field is parallel to the electric field, we
have $Y_{*}=E_{1}^{-2/3}$. According to (\ref{eq:45}), the creation
rate decreases when $\zeta_{T}$ increases. It coincides with our
result in bubble case. Moreover, the generic solution of (\ref{eq:46})
can be found as,

\begin{equation}
Y_{*}=\left[\frac{\vec{E}^{2}-\vec{B}^{2}+\sqrt{\left(\vec{E}^{2}-\vec{B}^{2}\right)^{2}+4\left(\vec{E}\cdot\vec{B}\right)^{2}}}{2}\right]^{-1/3}.
\end{equation}

\begin{figure}
\begin{centering}
\includegraphics[scale=0.4]{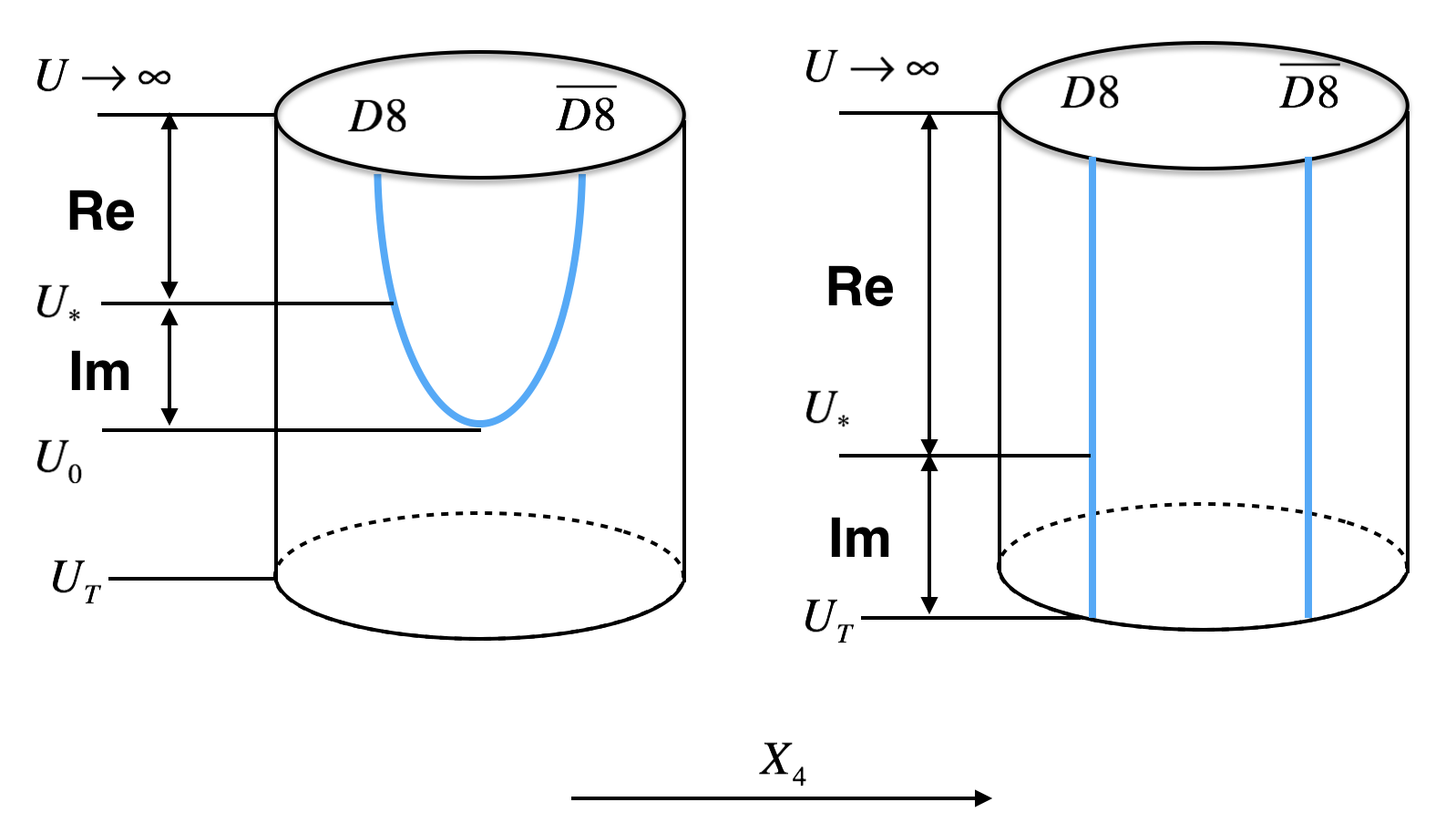}
\par\end{centering}
\caption{\label{fig:Figure 8} The region of the integral in the imaginary
Lagrangian of the D8/D8-branes in the black D0-D4 background. \textbf{Left:}
The configuration of D8/D8-branes is \textquotedblleft U\textquotedblright{}
shape, thus the integral in the imaginary Lagrangian starts from $U_{0}$
to $U_{\ast}$. \textbf{Right:} The configuration of D8/D8-branes
is parallel and the integral in the imaginary Lagrangian starts from
$U_{T}$ to $U_{\ast}$.}
\end{figure}

On the other hand, for the connected configuration of $\mathrm{D}8/\overline{\mathrm{D}8}$-branes,
taking the zero temperature limit $U_{T}\rightarrow0$, we have,

\begin{equation}
y_{0}=\frac{U_{T}}{U_{0}H_{T}H_{0}^{1/3}\left(U_{0}\right)}\rightarrow0.
\end{equation}
Thus the imaginary part of the Lagrangian contains the integral starting
from $Y_{0}$ to $Y_{*}$ (i.e. $U_{0}$ to $U_{*}$) with $Y_{0}\rightarrow Y_{*}$.
As the previous section, we can expand $Y$ (or $y$) in the neighborhood
$Y_{*}\simeq Y_{0}+\varepsilon$ where $\varepsilon\ll Y_{0}$. Since
in the connected configuration of the $\mathrm{D}8/\overline{\mathrm{D}8}$-branes,
$x_{4}$ is a function of $y$, we obtain the following behavior of
the imaginary Lagrangian in the small temperature limit $\chi\rightarrow0$,
\begin{align}
\mathrm{Im}\mathcal{L}\simeq & -\frac{2^{3}\pi^{2}\left(2\pi\alpha^{\prime}\right)^{7/3}R^{5}T_{8}}{3g_{s}}\vec{E}\cdot\vec{B}\left(1+\zeta_{T}\right)^{-1/2}\mathcal{G}\left(Y_{0}\right)\nonumber \\
 & \times\bigg\{1+\chi^{1/3}Y_{0}\left[1-Y_{0}^{3}\zeta_{T}\left(1+\zeta_{T}\right)^{3}\chi\right]^{4/3}\left[1-Y_{0}^{3}\left(1+\zeta_{T}\right)^{4}\chi\right]x_{4}^{\prime2}\left(Y_{0}\right)\bigg\}^{1/2}\varepsilon\nonumber \\
= & -\frac{2^{3}\pi^{2}\left(2\pi\alpha^{\prime}\right)^{7/3}R^{5}T_{8}}{3g_{s}}\vec{E}\cdot\vec{B}\left(1+\zeta_{T}\right)^{-1/2}\mathcal{G}\left(Y_{0}\right)\nonumber \\
 & \times\bigg\{\varepsilon^{2}+\chi^{1/3}Y_{0}\left[1-Y_{0}^{3}\zeta_{T}\left(1+\zeta_{T}\right)^{3}\chi\right]^{4/3}\left[1-Y_{0}^{3}\left(1+\zeta_{T}\right)^{4}\chi\right]\left[x_{4}\left(Y_{0}+\varepsilon\right)-x_{4}\left(Y_{0}\right)\right]^{2}\bigg\}^{1/2}=\mathrm{finite},
\end{align}
where

\begin{align}
\mathcal{G}\left(Y\right)= & Y^{-3/2}\left[1-Y^{3}\zeta_{T}\left(1+\zeta_{T}\right)^{3}\chi\right]^{-5/6}\nonumber \\
 & \left[1-Y^{3}\zeta_{T}\left(1+\zeta_{T}\right)^{4}\chi\right]^{-1/2}.
\end{align}
Therefore according to the evaluation, the imaginary part of the effective
Lagrangian is always finite at zero temperature limit. We have also
confirmed our conclusion numerically with arbitrary temperature. It
would be very interesting to compare our calculations in this section
with the D3/D7 approach in \cite{key-013,key-014}. The creation of
quark-antiquark is proportional to $\log\frac{1}{T}$ in \cite{key-013,key-014}
which diverges at zero temperature limit, while it is always finite
in our D0-D4/D8 system or the original Witten-Sakai-Sugimoto model.
However, this result is not surprised because there is a confining
scale in the compactified D4-brane (with or without smeared D0-branes)
system as (\ref{eq:7}) or (\ref{eq:8}), and on the other hand the
bubble configuration would be thermodynamically dominated at low temperature
while the black brane configuration arises at high temperature \cite{key-022,key-023,key-024,key-025}.
Consequently our calculation in the black D0-D4 configuration (high
temperature) consistently coincides with the case in bubble D0-D4
(low temperature) at zero temperature limit, which means the creation
rate should be definitely finite. Besides, as the situation of bubble
D0-D4 system, we also find an additionally possible instability based
on our calculations in (\ref{eq:43}) since the first line in (\ref{eq:43})
may also be imaginary. So it is the instability from the vacuum without
the electromagnetic field as discussed in the bubble case.

\section{Summary and discussion}

In this paper, we have studied the electromagnetic instability by
deriving the effective Euler-Heisenberg Lagrangian for the flavored
quarks in the Witten-Sakai-Sugimoto model with the D0-D4 background.
Since the dynamics of the flavored quarks is described by the DBI
action of the probe $\mathrm{D8/\overline{D8}}$-branes, we identify
its DBI action with the constant electromagnetic fields as the effective
Euler-Heisenberg action. Then we explore the electromagnetic instability
and evaluated the pair creation rate of quark-antiquark in the Schwinger
effect. With the D0-D4/D8 model in string theory, our investigation
contains the influence of the D0-brane density which could be interpreted
as the $\theta$ angle or chiral potential in QCD. In the bubble configuration,
since the D4-branes with smeared D0-branes are wrapped on a cycle,
it introduces a confining scale into this system. Therefore, we obtain
a very different result from $\mathcal{N}=2$ supersymmetric QCD in
the approach of D3/D7 \cite{key-013,key-014}.

In order to investigate the electromagnetic instability, we assume
the electromagnetic field is sufficient strong, then we find the $\theta$-dependent
creation rate of flavored quark-antiquark obtained in the bubble D0-D4
background exactly coincides with \cite{key-012} if setting $\theta$
angle or $\zeta=0$ (i.e. no D0-branes). Our numerical calculation
also shows the creation rate decreases when $\theta$ angle or $\zeta$
increases and its behavior depends on the direction of the magnetic
field relative to the electric field. To understand this, let us combine
our results with \cite{key-017,key-018,key-019,key-020}. Since the
critical electric field evaluated in (\ref{eq:37}) is in quantitative
agreement with the mass spectrum in \cite{key-017,key-018,key-019,key-020}
which describes the possible metastable states in the heavy-ion collision,
our results imply that in the heavy-ion collision the metastable state
could be created in the Schwinger effect then it soon decays to the
true vacuum as discussed in \cite{key-031,key-032}. And because of
the condensate of gluon, these metastable states become heavier, so
the critical electric field increases while its associated decay rate
decreases in the present of D0-branes i.e. $\theta$ angle.

Moreover, the creation rate in the black D0-D4 background has also
been computed which remains to be finite while it is oppose to the
D3/D7 approach \cite{key-013,key-014}. Nevertheless, our result would
be significant. Since the Hawking-Page transition in the WSS model
is usually interpreted as confined/deconfined phase transition in
QCD \cite{key-022,key-025}, it means the observables in the deconfined
phase should return to confined case if the temperature goes to zero.
So our results supports this statements qualitatively because it illustrates
the creation rate obtained in the black D0-D4 background (at finite
temperature) returns to the result from the bubble D0-D4 background
(zero temperature).

In addition, if turning off the electromagnetic fields, the effective
action (\ref{eq:39}) and (\ref{eq:43}) remains to include a vacuum
instability in the present of D0-branes. We suggest that this vacuum
instability might holographically describe the decay of the vacuum
with various winding numbers triggered by the $\theta$ angle or instantons
in QCD since the D0-branes relates to the $\theta$ angle thus could
be identified as instantons.

However, there might be some issues concerning the instability of
our holographic setup in this paper. For examples, first, our numerical
calculations are all based on the small $\theta$ angle or $\zeta$
expansion, so it is natural to ask what if we keep all the orders
of $\theta$ angle or $\zeta$? How the electromagnetic and vacuum
instability would be affected? Second, the electromagnetic field is
non-dynamical in our calculations. So what about a dynamical case?
Unfortunately, as opposed to the situation of the black D0-D4 background,
it seems impossible to introduce an electric current directly in the
bubble D0-D4 configuration since the flavor branes never end in the
bulk. To solve this problem, one may need a baryon vertex. But it
is less clear whether or not such a baryon vertex could be created.
We leave these issues to a future study.

\section*{ACKNOWLEDGMENT}

This work is inspired by our previous works \cite{key-018,key-019,key-020,key-021}
in USTC, and \cite{key-031} from our colleagues. We would like to
thank Chao Wu and Shi Pu for valuable comments and discussions. Wenhe
Cai is supported by the National Natural Science Foundation of China
under the Grant No. 11805117. Si-wen Li is supported by the research
startup foundation of Dalian Maritime University in 2019 and partially
by the National Natural Science Foundation of China under the Grant
No. 11535012.

\end{document}